\documentclass{emulateapj} 

\usepackage{apjfonts}
\usepackage{natbib}

\shorttitle{SPATIALLY RESOLVED GAS KINEMATICS IN AGN} 
\shortauthors{RICE ET AL.}

\newcommand{\mbh}{M_\bullet}
\newcommand{\hst}{{\it HST}}
\newcommand{\kms}{km s$^{-1}$}
\newcommand{\oiii}{[\ion{O}{3}]}
\newcommand{\oii}{[\ion{O}{2}]}

\newcommand{\sii}{[\ion{S}{2}]}

\slugcomment{ApJ accepted [15 Sept 2005]}

\begin{document}

\title{Spatially-Resolved Narrow Line Region Kinematics in 
Active Galactic Nuclei} 

\author{Melissa S. Rice, Paul Martini\altaffilmark{1}, Jenny E. Greene} 

\affil{Harvard-Smithsonian Center for Astrophysics; 
60 Garden Street, MS20; Cambridge, MA 02138} 

\altaffiltext{1}{Current Address: Department of Astronomy; The Ohio State 
University; 140 West 18th Avenue; Columbus, OH 43210; 
martini@astronomy.ohio-state.edu} 

\author{Richard W. Pogge}
\affil{Department of Astronomy; The Ohio State University; 140 West 18th 
Avenue; Columbus, OH 43210}

\author{Joseph C. Shields} 
\affil{Physics \& Astronomy Department; Ohio University; Athens, OH 45701} 

\author{John S. Mulchaey}
\affil{Carnegie Observatories; 813 Santa Barbara Street; Pasadena, CA 91101}

\and

\author{Michael W. Regan}
\affil{Space Telescope Science Institute; 3700 San Martin Drive;
Baltimore, MD 21218}

\begin{abstract}

We have analyzed {\it Hubble Space Telescope} spectroscopy of 24 nearby 
Active Galactic Nuclei (AGNs) to investigate spatially-resolved gas kinematics 
in the Narrow Line Region (NLR). 
These observations effectively isolate the nuclear line profiles on less than 
100 parsec scales and are used to investigate the origin of the substantial 
scatter between the widths of strong NLR lines and the stellar velocity 
dispersion $\sigma_*$ of the host galaxy, a quantity which relates with 
substantially less scatter to the mass of the central, supermassive black hole, 
and more generally characterize variations in the NLR velocity field 
with radius.  
We find that line widths measured with STIS at a range of spatial scales 
systematically underestimate both $\sigma_*$ and the line width measured from 
ground-based observations, although they do have comparably large scatter to 
the relation between ground-based NLR line width and $\sigma_*$. 
There are no obvious trends in the residuals when compared with a range of 
host galaxy and nuclear properties. 
The widths and asymmetries of \oiii\ $\lambda5007$ and 
\sii\ $\lambda\lambda6716, 6731$ as a function of radius exhibit a 
wide range of behavior. Some of the most common phenomena are 
substantial width increases from the STIS to the large-scale, ground-based 
aperture and almost no change in line profile between the unresolved nuclear 
spectrum and ground-based measurements. 
We identify asymmetries in a surprisingly large fraction of low-ionization 
\sii\ line profiles and several examples of substantial red asymmetries in 
both \oiii\ and \sii. 
These results underscore the complexity of the circumnuclear material 
that constitutes the NLR and suggests that the scatter in the NLR width and 
$\sigma_*$ correlation can not be substantially reduced with a simple set 
of empirical relations. 

\end{abstract}


\section{Introduction} \label{sec:intro}

The width of emission lines from the Narrow Line Region (NLR) in Active 
Galactic Nuclei (AGNs) has recently received a great deal of interest because 
it may provide a reasonably accurate, although not precise, estimate of the 
host galaxy spheroid velocity dispersion $\sigma_*$. 
The NLR $\sigma_g =$ FWHM/2.354 may therefore be a reasonable `tertiary' black 
hole mass $\mbh$ estimator based on a series of empirical relations that 
originate with the $\mbh-\sigma_*$ relationship between the mass of a 
galaxy's central, supermassive black hole and the stellar velocity dispersion 
of the host galaxy's spheroid \citep{Ferrarese00,Gebhardt00a} and 
the relation between $\sigma_*$ and NLR line width from \citet[][and earlier 
work by Whittle 1992b,c]{Nelson96}. 
Subsequent measurements and analysis have found that the $\mbh-\sigma_*$ 
relationship has an intrinsic scatter of no more than 0.3 dex in ${\rm log} 
\mbh$ and that this upper limit is due to present measurement uncertainties 
\citep{Tremaine02}. \citet{Nelson00a} used this relation and 
$\mbh$ estimates from reverberation-mapping experiments to propose the use of 
$\sigma_g$ measured from the \oiii\ $\lambda5007$ line to estimate 
the black hole mass. 
\citet{Greene05} recently completed a direct comparison of $\sigma_*$ and the 
gas velocity dispersion $\sigma_g$ for approximately 2000 AGNs from the 
Sloan Digital Sky Survey (SDSS) and showed that these widths are well 
correlated, although with considerable scatter. 

The empirical relation between $\sigma_g$ and $\mbh$ is important for 
a broad range of applications. These applications include estimates of 
$\mbh$ for local AGNs, for AGNs at high redshift, and the cosmic evolution 
of supermassive black holes. For luminous AGNs with substantial continuum 
emission, or for AGNs at high redshift, the widths of the narrow emission 
lines may be the only method to determine $\sigma_*$. In conjunction with 
measurements of bolometric luminosity $L_{bol}$, estimates of $\mbh$ 
can also be used to calculate the accretion luminosity of AGN in terms of 
the Eddington ratio $L_{bol}/L_{Edd}$. At low-redshifts, black hole 
estimates from reverberation mapping have been shown to agree quite well 
with the slope of the slope of the $\mbh-\sigma_*$ relation
\citep{Gebhardt00b,Ferrarese01,Nelson04,Onken04} and have been used to 
calibrate an additional secondary (virial) $\mbh$ estimator based on the 
line width and luminosity of the broad, permitted lines 
\citep[e.g.][]{Kaspi00}. 
\citet{Boroson03} used this virial relation to show that the \oiii\ FWHM could 
be used to estimate $\mbh$ to within a factor of five. The virial and 
$\sigma_g$ estimates of $\mbh$ have been used to explore the black hole 
masses and accretion rates in particular classes of AGNs, most notably the 
Narrow Line Seyfert 1s \citep{Grupe04}, as well as to search for evolution 
in the $\mbh-\sigma_*$ relation \citep{Shields03}. Measurements of 
evolution in the $\mbh-\sigma_*$ relation, or skewness in the distribution 
of the scatter about this relation, could also provide valuable constraints on 
the formation history of supermassive black holes \citep{Robertson05}. 

The utility of $\sigma_g$ measurements as a proxy for $\mbh$ in AGNs is largely 
limited by the substantial scatter in the relation between $\sigma_*$ and 
$\sigma_g$. The brightest line for $\sigma_g$ measurements is the 
\oiii\ $\lambda 5007$ line, although this line does suffer from substantial 
blue-side asymmetries \citep[e.g.][]{Heckman81} that dramatically 
affect the quality of the $\sigma_g - \sigma_*$ correlation. 
\citet{Greene05} found that $\sigma_g/\sigma_* = 1.34 \pm 0.66$ for 
the full profile of the \oiii\ $\lambda 5007$, while after removal of the 
blue-side asymmetry the result for the line core is 
$\sigma_g/\sigma_* = 1.00 \pm 0.35$. The other bright NLR lines, 
\oii\ $\lambda3727$ and \sii\ $\lambda\lambda6716, 6731$, have lower 
ionization potentials and do not suffer from such substantial asymmetries. 
The scatter and quality of the $\sigma_g - \sigma_*$ relation for these 
lines is comparable to the core of the \oiii\ line. 
While all of these correlations do strongly confirm the \citet{Nelson96} 
result that the kinematics of the NLR are dominated by gravity, the origin of 
the scatter is not clear. 
One potential origin of this scatter is a nonvirial contribution
to the emission line widths. These contributions are clearly present in 
the blue-asymmetric \oiii\ lines, but because 
these asymmetries are significantly stronger in high ionization lines like 
\oiii, compared to lower ionization lines like \sii\ 
\citep[e.g.][]{deRobertis84}, 
there is good evidence that a range of spatial scales contribute to the 
observed NLR kinematics. 
Beyond the gas kinematics that contribute to the \oiii\ blue wings, 
any nonvirial contribution to deviations from the $\sigma_g - \sigma_*$ 
correlation is evidently traced equally by the low ionization \sii\ lines 
and the core of the higher ionization \oiii\ line, since these features 
show similar scatter. 

Other potential origins for the scatter in the $\sigma_g - \sigma_*$ 
relation include the impact of compact radio jets and tidal distortions of the 
host galaxies. \citet{Nelson96} found that both of these quantities 
correlate with systematically larger $\sigma_g$ at a given $\sigma_*$, 
although large-scale radio jets or tidal distortions are not present in most 
galaxies and therefore can not be responsible for the bulk of the scatter. 
Small-scale radio jets may play a more significant role, however, as 
\citet{Ho01} find approximately 60\% of Seyfert 1 galaxies qualify as 
radio loud when the nuclear radio and visible-wavelength flux is isolated 
from host galaxy emission. \citet{Whittle92c} showed that the \oiii\ lines 
are significantly broader in Seyferts with linear radio morphology and high 
radio luminosity, which suggests radio jets can affect the NLR velocity 
field. 
The spatial distribution of the NLR gas within the bulge may also have 
a substantial impact on the observed value of $\sigma_g$. While the 
NLR gas must approximately trace the spheroid kinematics to produce 
this correlation, spatially-resolved images indicate that the NLR is often 
approximately confined to a plane \citep[e.g.][]{Pogge89b}. \citet{Whittle92b} 
found that rotation does contribute to the width of the \oiii\ emission line, 
although it does not dominate. Finally, \citet{Greene05} investigated if host 
galaxy morphology, local environment, star formation rate, AGN luminosity, 
and Eddington ratio correlated with the observed scatter and only found 
evidence for systematically larger $\sigma_g$ at higher Eddington ratio. 
While all of these investigations have determined that tidal 
disturbances, the presence of a radio source, rotation, and Eddington ratio 
contribute to the observed scatter, none of them dominate. 
It would be extremely valuable if the origin of the scatter in the 
$\sigma_g - \sigma_*$ correlation could be identified and substantially 
reduced, as then $\sigma_*$ (and $\mbh$) could be estimated more precisely 
for AGNs. 

Further information on the origin of the scatter in the $\sigma_g - \sigma_*$ 
correlation, as well as a more general understanding of the NLR kinematics, 
may be gained from a high spatial resolution study. 
Nearly all studies of NLR kinematics to date have employed ground-based, 
single-aperture measurements with aperture sizes on the order of a few 
arcseconds. Such aperture sizes correspond to several hundred parsecs in 
projection and can include most of the NLR. There is good evidence, however, 
for the presence of stratification in the NLR, such as observations that 
higher-ionization and higher critical density lines tend to be broader 
\citep{deRobertis86}, the presence of a blue, asymmetric wing on the 
higher-ionization \oiii\ line and not on other, lower-ionization lines 
such as \sii, and evidence for modest increases in electron density from 
recent \hst\ observations \citep{Barth01,Shields05}. 
The properties of \oiii\ led \citet{Heckman81} to propose that asymmetries in 
the NLR are due to radial outflow and 
wind models of the NLR \citep[e.g.][]{Krolik84,Schiano86,Smith93} are supported 
by recent, spatially-resolved \hst\ STIS kinematics of NGC~4151 
\citep{Nelson00b,Das05}. 
These authors find evidence for both a rotational component and two, 
kinematically-distinct radial outflow components that appear to decelerate at 
larger scales. 
\citet{Crenshaw00} similarly studied the central 400 parsecs of NGC~1068 and 
found evidence for radial outflow driven by either wind or radiation pressure 
in the nuclear region, followed by deceleration on larger spatial scales. 

In the present paper we analyze spatially-resolved spectroscopy of a large
sample of nearby AGN from archival STIS observations and 
measure the NLR kinematics as a function of aperture size. We investigate 
the origin of the scatter in the $\sigma_g - \sigma_*$ correlation 
through a range of line width measurements at increasing aperture size, 
ranging from the limits of \hst\ resolution to an aperture size typical 
of ground-based studies. We also study line asymmetries as a function of 
aperture size to investigate the spatial origin of line 
asymmetries. The sample properties and selection 
are described next in \S\ref{sec:sample} and the nonstandard aspects of 
the data processing in \S\ref{sec:datar}. We describe the 
line width and asymmetry parameters in \S\ref{sec:lineprof}, along with 
the results of these measurements for different spatial scales in 
\S\ref{sec:res}. The implications of this work for the $\sigma_g - \sigma_*$ 
correlation and the kinematics of the NLR are then presented in 
\S\ref{sec:dis}. We summarize our results in the final section.

\section{The Sample}  \label{sec:sample}

We have adopted the sample of \citet[][hereafter NW95]{Nelson95} as the parent 
sample of this investigation of small-scale NLR kinematics because they have
obtained high-quality ground-based measurements of many NLR emission features, 
as well as stellar velocity dispersions, for a wide range of galaxy
types.  Their sample is largely drawn from the 140 Seyfert
galaxies discussed in \citet{Whittle92a}, where their primary selection
criteria was for strong Mg {\it b} ($\sim$5175\AA) and \ion{Ca}{2} triplet 
($\sim$8550\AA) absorption features and weak \ion{Fe}{2} emission (which 
confuses Mg {\it b}). With moderate spectral resolution (80--230 \kms\ FWHM) 
they measured the redshift, stellar kinematics ($V_{*}$ and $\sigma_{*}$),
and a variety of characteristics for the \oiii\ $\lambda5007$ and 
\sii\ $\lambda9096$ lines (see \S\ref{sec:lineprof}). 
These high-quality ground-based data are a critical point of comparison for 
our study of the small-scale kinematics. 

We searched the \hst\ archive and found 24 AGNs from the NW95 sample
with moderate-resolution spectroscopy of either the \oiii\ or
\sii\ emission lines (G430M and G750M gratings, respectively) through 
either an $0.1''$ or $0.2''$ wide slit. 
The spectral resolution of the G750L and G430L gratings ($\sim$500 \kms) 
is not high enough to resolve line width changes in nearly
all of these sources, while in contrast the instrumental resolution is 
substantially higher for G430M and G750M. 
The $0.1''$ and $0.2''$ slit widths have a spectral resolution of 
$\sim 1.5$ pixels FWHM for point sources and the G430M and G750M gratings 
have dispersions of 0.28 and 0.56 \AA/pixel. The instrumental velocity 
resolution FWHM is therefore 25 and 40 \kms, respectively.  
The resolution is approximately a factor of $\sim 2.7$ times larger for an 
unresolved source that uniformly fills the $0.2''$ slit and a factor of 
$\sim 1.7$ times larger for the $0.1''$ slit.
The STIS plate scale is $0.05078''$ per pixel and the slit length is $50''$. 
The details of our sample, including host galaxy properties, program ID, and 
observing mode, are provided in Table~\ref{tbl:data}. 
Of the 24 galaxies, only NGC~2110 has observations with both the G430M and 
G750M settings. 

\begin{deluxetable*}{lccccccccccccc}
\tablenum{1}
\tabletypesize{\tiny}
\tablewidth{0pt}
\tablecolumns{14}
\tablecaption{Observation and Sample Properties\label{tbl:data}}
\tablehead{
\colhead{Name} &
\colhead{Class} &
\colhead{cz} &
\colhead{Dist.} &
\colhead{$D$} &
\colhead{Morph.} &
\colhead{log$L_{1.4GHz}$} &
\colhead{ID} &  
\colhead{Mode} &
\colhead{CR-SPLIT} &
\colhead{Apertures} &
\colhead{Refs.} &
\colhead{Refs.} & 
\colhead{Refs.} \\
\colhead{} &
\colhead{} &
\colhead{(\kms)} &
\colhead{(Mpc)} &
\colhead{('')}&
\colhead{(T)} &
\colhead{(W/Hz)} &
\colhead{} &
\colhead{} &
\colhead{} &
\colhead{('')} &
\colhead{Dist.} &
\colhead{$D$} &
\colhead{Morph.} \\
\colhead{(1)} &
\colhead{(2)} &
\colhead{(3)} &
\colhead{(4)}&
\colhead{(5)} &
\colhead{(6)} &
\colhead{(7)} &
\colhead{(8)} &
\colhead{(9)} &
\colhead{(10)} &
\colhead{(11)} &
\colhead{(12)} &
\colhead{(13)} &
\colhead{(14)} \\
}
\startdata
MKN 270  & 2 & 3137  & 38.1 & 65.8 & -2 & 21.51 &  9143 & G750M & 1 & 52X0.2    & 1& 1 & 3 \\
MKN 573  & 2 &   5166  &  68.9 & 101.9 & -1 & 21.56 &  9143 & G750M & 1 & 52X0.2& 4 & 2 & 3   \\
MKN 686  & 2 &  4251   & 56.7 & 90.8 & 5 & 22.51 &  9143 & G750M & 1 & 52X0.2   & 4 & 2 & 1   \\
NGC 788  & 2 &   4061  &  54.1 & 77.3 & 0 & - - - &  9143 & G750M & 1 & 52X0.2  & 4 & 2 & 3   \\
NGC 1052 & 2 &  1307  &  18.0 & 77.3 & -5 & - - - &  7403 & G750M & 3 & 52X0.2  & 2 & 1 & 3   \\
NGC 1358 & 2 &   4008   & 53.4 & 119.7 & 0 & - - - &  9143 & G750M & 1 & 52X0.2 & 4 & 1 & 3   \\
NGC 1667 & 2 &  4547 & 60.6 & 53.5 & 5 & - - - &  9143 & G750M & 1 & 52X0.2 & 2 & 1 & 3   \\
NGC 2110 & 2 &   2249   & 30.0 & 42.5 & -3 & - - - &  8610 & G750M & 6 & 52X0.2 & 4 & 2 & 3   \\
NGC 2273 & 2 &  1929   & 27.6 & 203.3 & 5 & 21.84 &  9143 & G750M & 1 & 52X0.2  & 1 & 1 & 1   \\
NGC 3031 & 1.8 &   -39  &  1.4 & 1439.3 & 3 & 20.81 &  7351 & G750M & 2 & 52X0.1& 1 & 1 & 3    \\
NGC 3227 & 1.5 &  1024  &  20.6 & 119.7 & 1 & 21.84 &  7403 & G750M & 2 & 52X0.2& 1 & 1 & 3   \\
NGC 3516 & 1.5 &  2503  &  38.9 & 137.5 & -2 & 21.85 &  8055 & G750M & 1&52X0.2 & 1 & 1 & 3 \\
NGC 3982 & 2 &  1155  &  17.0 & 109.2 & 3 & 21.54 &  7361 & G750M & 1 & 52X0.2  & 1 & 1 & 3 \\
NGC 4051 & 1.5 &   622  &  17.0 & 361.5 & 4 & 21.35 &  8228 & G750M & 1 & 52X0.2& 1 & 1 & 3 \\
NGC 4579 & 1.9 & 1334 &  19.1 & 378.6 & 3 & 21.34  &  7403 & G750M & 3 & 52X0.2 & 3 & 1 & 3 \\
NGC 5347 & 2 &  2295  & 31.1 & 101.9 & 2 & 21.07 &  9143 & G750M & 1 & 52X0.2   & 1 & 1 & 3 \\
NGC 5427 & 2 &  2733  & 33.2 & 181.2 & 5 & - - - &  9143 & G750M & 1 & 52X0.2   & 1 & 1 & 3 \\
NGC 7682 & 2 &  5109  & 68.1 & 65.8 & 2 & 22.59 &  9143 & G750M & 1 & 52X0.2    & 4 & 1 & 3 \\
 & & & & & & & & & & & &  & \\
MKN 348  & 2 &   4505  & 60.1 & 75.5 & 0 & 22.17  &  8253 & G430M & 2 & 52X0.2  & 4 & 2 & 3  \\
MKN 1066 & 2 &  3523   & 47.0 & 72.1 & -5 & 23.16 &  8253 & G430M & 2 & 52X0.2  & 4 & 2 & 3   \\
NGC 2110 & 2 &   2249   & 30.0 & 42.5 & -3 & - - - &  8253 & G430M & 2 & 52X0.2 & 4 & 2 & 3   \\
NGC 4151 & 1.5 &  966  &  20.3 & 424.8 & 2 & 22.25 &  8473 & G430M & 3 & 52X0.2 & 1 & 1 & 3  \\
NGC 5194 & 2   &   461 &  7.7 & 534.8 & 4 & 22.20 &  7574 & G430M & 2 & 52X0.2  & 1 & 1 & 3  \\
NGC 5929 & 2 &  2250 & 38.5 & 65.8 & 2 & 22.22  &  8253 & G430M & 2 & 52X0.2    & 1 & 1 & 3  \\
NGC 7674 & 2 &  8673 & 115.6 & 60.0 & 4 & 23.63 &  8259 & G430M & 2 & 52X0.2    & 4 & 1 & 3  \\
\enddata
\tablecomments{
Col. (1): Galaxy name. Several common aliases are MKN~270 = NGC~5283, 
MKN~686 = NGC~5695, NGC~3031 = M81, and NGC~5194 = M51a. Col. (2): AGN type, 
where Type 1 AGN exhibit broad permitted lines and Type 2 AGN have narrow 
permitted lines. 
Col. (3): [O~III] profile median redshift from \citet{Nelson95} ([O~III] 
C80 or [S~III] C80 redshifts used for galaxies with no [O~III] median 
redshift given). Col. (4): Distance. Col. (5): Angular diameter in arcsec 
of the 25 mag arcsec$^{-2}$ isophote. Col. (6): Morphological 
type (de Vaucouleurs numerical type). Col. (7): Radio luminosity (log$_{10}$
of the absolute spectral luminosity at 1.4GHz) from \citet{Condon02}. 
Col. (8): \hst\ proposal ID for data used. Col. (9): STIS grating. 
Col. (10): Number of CR-split exposures. Col. (11): STIS aperture. 
Cols. (12)--(14): References to published distances, sizes and morphologies.
}
\tablerefs{Distances: 
(1) \citet{Tully88}; (2) \citet{Jensen03}; 
(3) \citet{Solanes02}; (4) $H_0 = 75$ \kms\ Mpc$^{-1}$ (adopted for 
consistency with \citet{Tully88}). 
Size and Morphology: 
(1) \citet{Nilson73}; (2) \citet{VV74}; (3) \citet{deVaucouleurs91} \\
}
\end{deluxetable*}

\section{Data Reduction}  \label{sec:datar}

Our sample is drawn from many different STIS observing programs and thus the 
objects were obtained with many different observing sequences. 
The reduction steps therefore vary from galaxy to galaxy depending upon the 
number of dithered exposures, the number of CR-split frames, and if the 
observations were obtained prior to the primary (Side 1) STIS electronics 
failure on 2001 May 16. 
We found that the reduced spectra taken directly out of the CALSTIS pipeline 
are not of optimum quality for any of our galaxies. In this section we discuss 
our modifications to the CALSTIS pipeline for the various observing sequences 
employed to obtain these archival data.

\subsection{\it Multiple Dithered Exposures} \label{sec:dither}

Twelve of our 24 galaxies (those obtained for proposals 7361, 8055, and 9143) 
include multiple exposures of the nucleus and employed integer pixel 
dithering along the slit. These exposures had only one CR-split frame, and 
thus the CALSTIS pipeline reduction did not include cosmic ray correction. 
To account for the cosmic rays, we broke out of the CALSTIS pipeline after 
the dark-subtraction, bias-subtraction, and flat-fielding for each
exposure (using the \_flt.fits calibration file). 
After aligning the exposures, we used the Laplacian Cosmic Ray identification 
IRAF routine \citep[L.A. Cosmic;][]{vanDokkum01} to create a bad pixel map for 
each 2-D spectrum. The pixels flagged as cosmic rays by L.A. Cosmic were then 
rejected when the multiple spectra were combined.
L.A. Cosmic occasionally mistook spectral features for cosmic rays, leaving 
emission lines blank in the final combined image. 
In these cases we replaced the empty pixels with the corresponding values 
from the original \_flt image.  
Once the cosmic-ray cleaning was complete, we re-inserted this combined, 
cosmic ray-rejected file into the CALSTIS pipeline for the final wavelength 
and flux calibrations.  

\subsection{\it Multiple CR-splits} \label{sec:crsplit}

When multiple images were obtained at the same pointing, the exposures could 
be combined with the {\tt ocrreject} step in the CALSTIS pipeline to produce 
a cosmic ray corrected image.  We have run the spectra from galaxies with 
multiple CR-splits listed in Table~\ref{tbl:data} up through this cosmic ray 
rejection step in the CALSTIS pipeline.  However, even with multiple 
CR-splits, the pipeline output most often is still peppered with cosmic rays.  
We thus used L.A. Cosmic to zap the remaining rays in the ocrreject output 
(\_crj.fits image) before re-inserting the spectra into the pipeline for the 
wavelength and flux calibrations.

\subsection{\it Side 2 STIS Electronics Correction} \label{sec:side2}

The ten datasets for proposal 9143 were obtained after 2001 May 16, when the
Side 1 STIS electronics failed. 
The Side 2 electronics do not have closed-loop temperature control
of the CCD and this results in a dark rate that varies with temperature. 
Thus the dark calibration image is a poor approximation to the real dark 
rates for hot pixels and the outputs of the CALSTIS pipeline have many 
strongly negative pixels after the dark-subtraction step. 
Unfortunately, even the dithering scheme employed for this proposal did not 
easily remove the negative 'holes' created in the images.

To correct for these negative pixels, we roughly followed the method outlined
by Pogge et al. (2005, {\it in preparation}). 
From the combined, cosmic ray-rejected image for each galaxy (see 
\S\ref{sec:dither}), we created a list of pixels that fell more 
than 3-$\sigma$ below the background median, which we fed into a custom IRAF 
script that replaced each negative pixel with the mean of its surrounding 
3$\times$3-pixel box. We then re-inserted this corrected image into the 
CALSTIS pipeline. For images obtained prior to the Side 1 electronics failure,  
hot pixels that remained after the dark correction were eliminated as 
part of the cosmic ray cleaning step. 

\section{Data Analysis}  \label{sec:lineprof}

Once these steps were complete, we extracted 1-D spectra of different 
aperture sizes from the 2-D spectra using the CALSTIS {\tt x1d} routine, 
which performs a geometric rectification and background subtraction. 
For each galaxy we extracted spectra with apertures of integer pixel sizes 
from 2--50 centered on the nucleus, corresponding to angular sizes of 
$0.10'' - 2.54''$. We chose to integrate progressively wider 
apertures, rather than attempt to measure the differential line profiles 
as a function of radius, because one of the main goals of our study is to 
characterize the transition between NLR profiles on very small scales and 
the ground-based apertures used in previous work. 

Figure~\ref{fig:spectra} presents three 1-D spectra for each of the 24 
galaxies in the sample. 
The leftmost column of panels plots each galaxy's spectrum summed over an 
$0.2''$ aperture, which approximately represents the nuclear emission 
unresolved by STIS. The middle column of panels plots the sum of a $1''$ 
aperture centered on the nucleus. The $1''$ aperture was chosen as a compromise 
between the resolution of previous ground-based studies and the angular extent 
over which emission lines are detectable in most of the STIS spectra. 
Finally, the rightmost column of panels shows the difference of the 
middle and leftmost panels. The difference spectra in these panels thus 
illustrate the line profiles {\it outside} of the nucleus and will be 
discussed below in the context of line widths and asymmetries outside 
of the nuclear region.  

\begin{figure*}
\figurenum{1}
\plotone{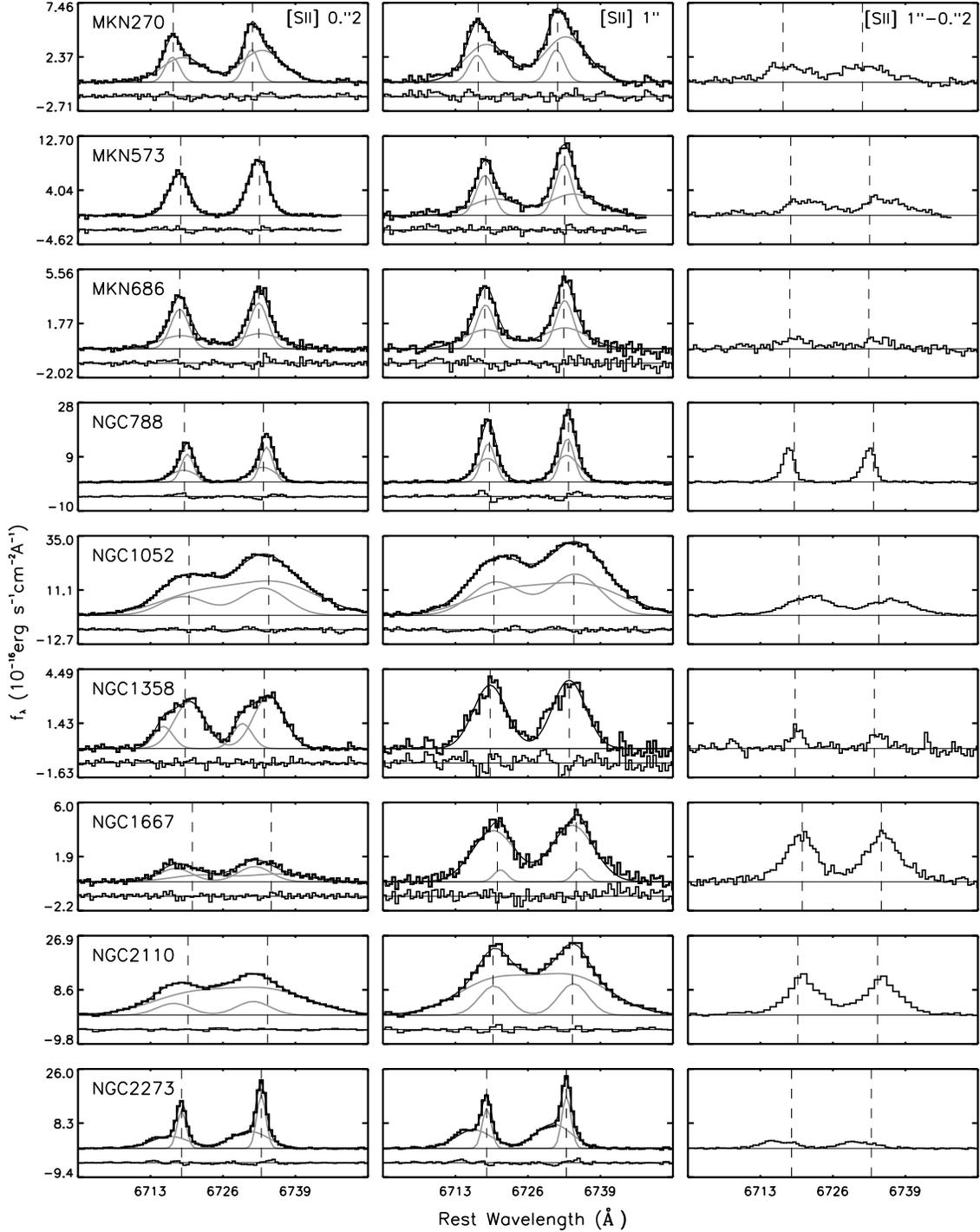}
\caption{Profiles for the \sii\ $\lambda\lambda$6717,6731
or \oiii\ $\lambda$5007 lines for our sample are
presented at aperture sizes of $0.2''$ (left column) and $1''$ (middle
column), as well as for $1''-0.2''$ (right column).  
The best fit Gaussian profiles are shown as smooth, solid lines, while
the individual components are shown as grey lines. 
Residuals are plotted below each fit. 
The vertical dashed line represents the position of the centroid of each line 
profile in the $1''$ aperture.  A polynomial fit to the AGN 
continuum has been subtracted from NGC 3227, NGC 3516, NGC 4051, and
NGC 4579. All other profiles have only had a constant subtracted.
The profiles have been plotted in rest wavelengths using the cz values
from in Table~\ref{tbl:data}.
Note that a Gaussian profile was not fit to NGC 7682 because the STIS G750M 
grating was centered too far to the blue and the broad, red wing of the 
\sii\ $\lambda 6731$ line did not fall on the detector.} \label{fig:spectra} 
\end{figure*}

\begin{figure*}
\figurenum{1}
\plotone{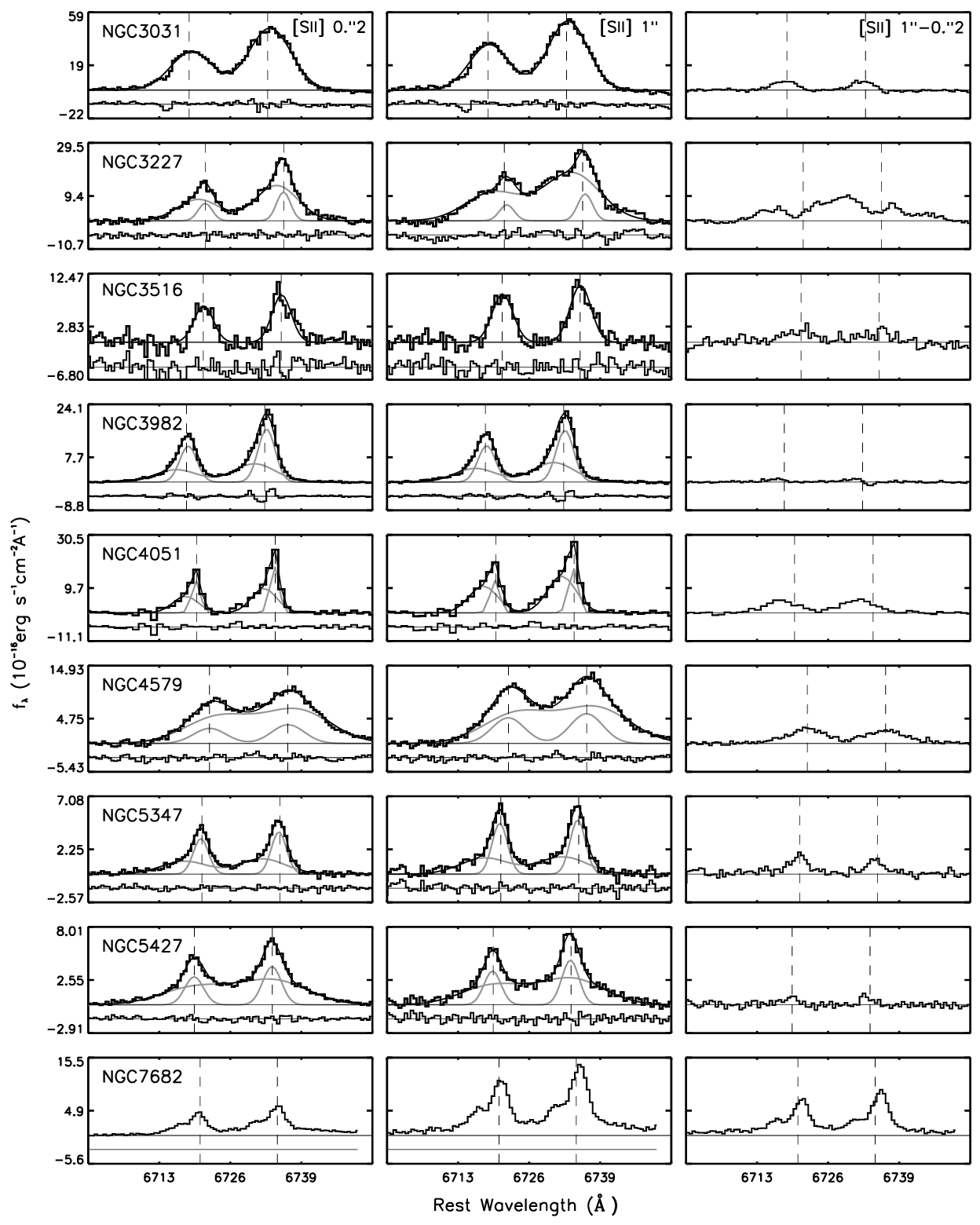} 
\caption{\it Continued}
\end{figure*}

\begin{figure*}
\figurenum{1}
\plotone{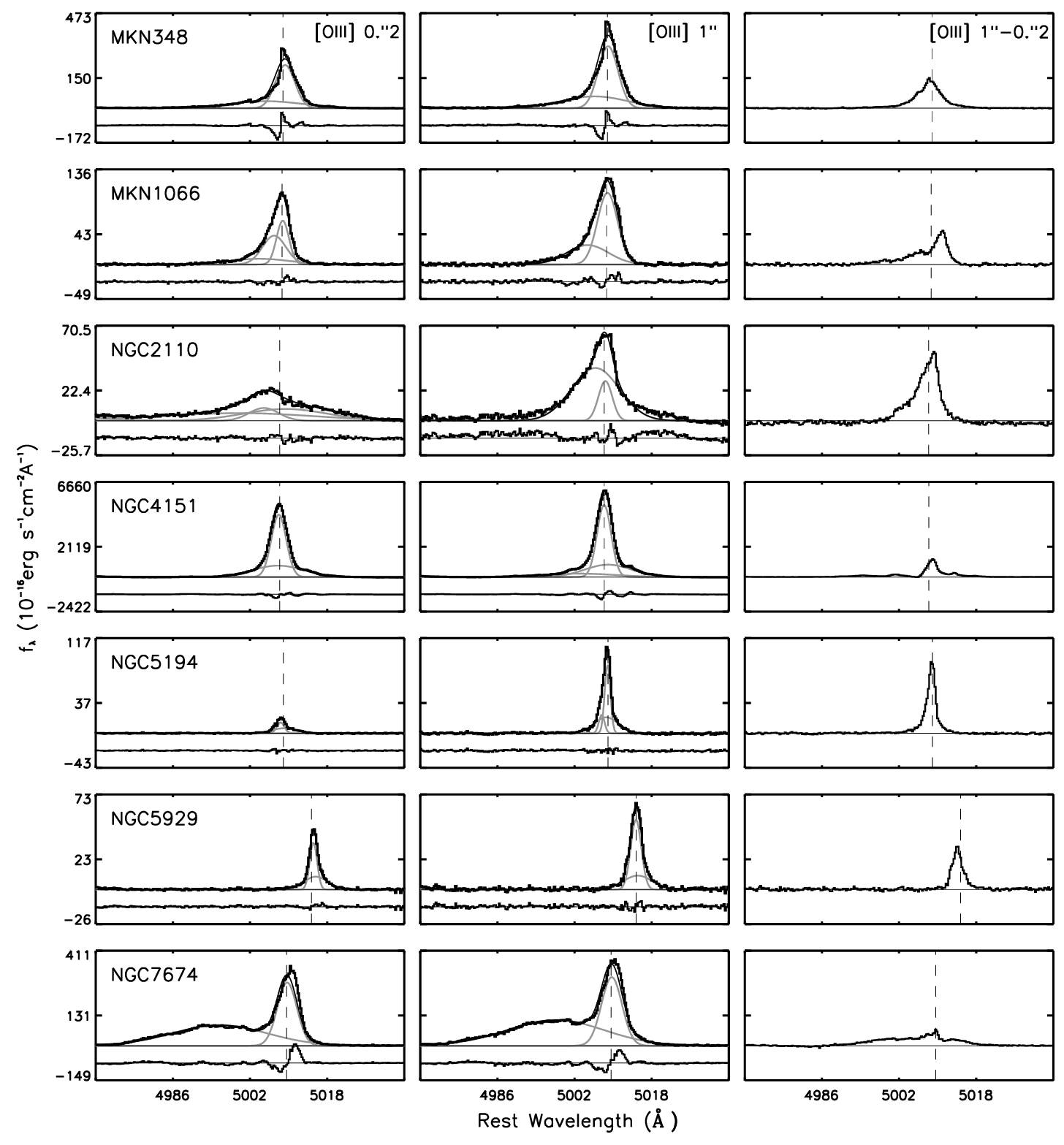} 
\caption{\it Continued}
\end{figure*}

We applied a scalar throughput correction to the flux-calibrated $0.2''$
spectra (4-pixel extraction box height) produced by x1d because the default
throughput correction table (referenced by the PCTAB keyword) does not include 
an entry for a 4-pixel extraction box height. 
When a 4-pixel extraction box height is requested, the throughput value for 
a 3-pixel box height is employed instead by default, although the throughput 
is approximately 10\% larger for a 4-pixel box compared to a 3-pixel box. 
The default behavior of x1d therefore overestimates the flux in a 4-pixel 
box when it assumes the lower throughput appropriate for a 3-pixel box. 
To properly correct the $0.2''$ spectra, we interpolated the throughput 
values in the correction table to calculate the appropriate quantity for 
a 4-pixel box height for each instrument configuration and then applied this 
throughput correction to the x1d output. 
Also, we note that while the throughput correction in the table was calculated 
from observations of a point source, the AGN in this sample are dominated 
by unresolved nuclear emission and this correction table is a reasonable 
approximation. 
This throughput correction is only relevant to the $1'' - 0.2''$ and 
$0.2''$ spectra shown in Figure~\ref{fig:spectra} because the correction is 
negligible at $1''$. The correction also does not affect the width 
measurements because the spectra are only multipled by a constant.

\begin{figure*}
\figurenum{2}
\label{fig:fig2}
\plotone{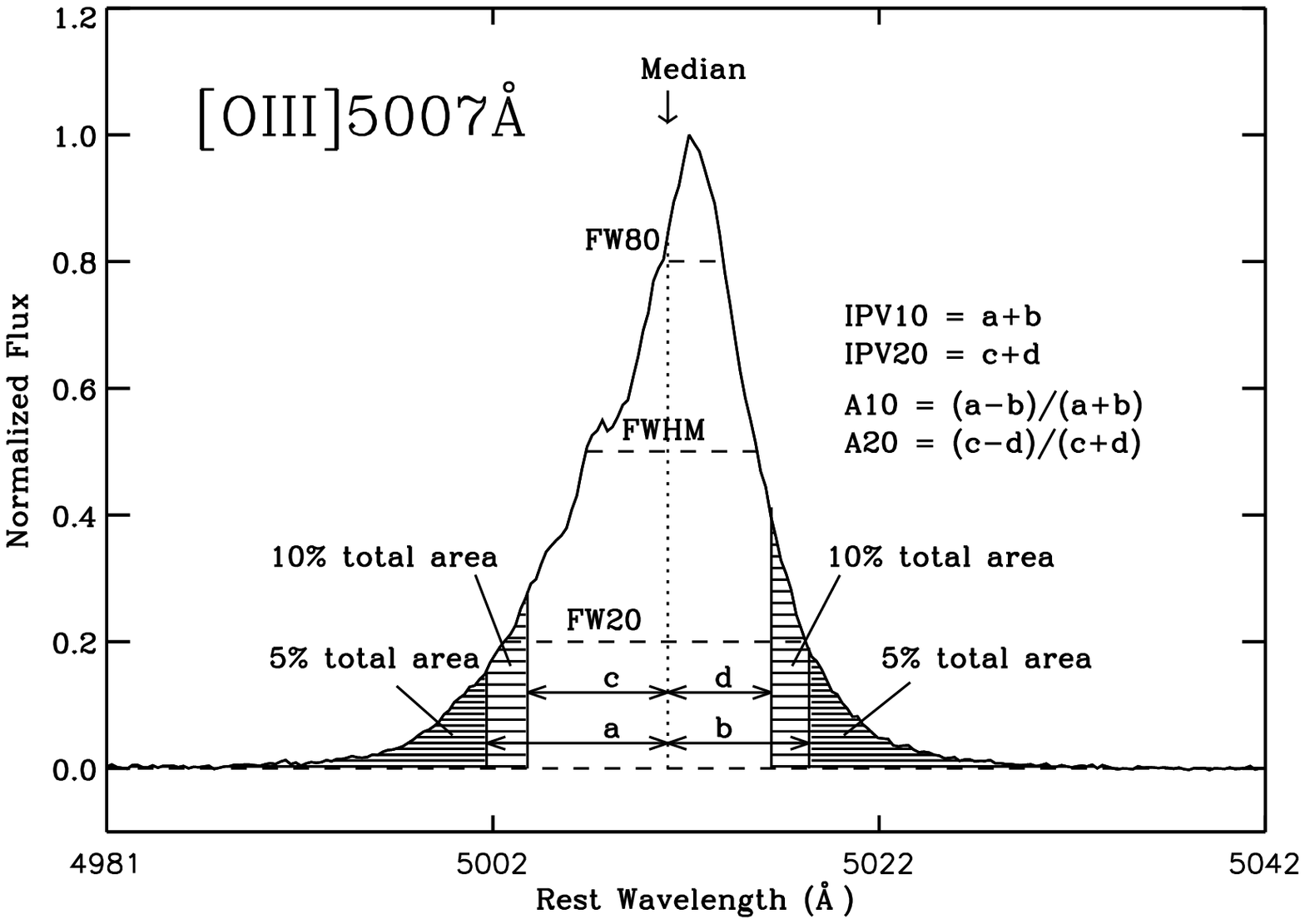}
\caption{A sample \oiii\ $\lambda$ 5007 profile ($0.2''$ aperture) with
definitions of the width parameters FW20,
FWHM, FW80, IPV10, and IPV20, and the asymmetry parameters A10 and
A20.  Note the broad wings and blue asymmetry typically seen in
AGN \oiii\ profiles.} \label{fig:def}
\end{figure*}

We have chosen to characterize the line profiles of the \oiii\ and \sii\ 
emission lines as a function of aperture size through line widths 
measured at a range of heights, interpercentile velocity widths (IPVs), and 
asymmetry measures \citep[e.g.][]{Heckman81,Whittle85a}. 
The definitions of these parameters, our measurement procedure, and their 
physical motivation, are discussed in detail below.  
The measurements of each of these quantities are listed in 
Tables 3 and 4 for $0.2''$ and $1''$ apertures. 
We present an illustration of these parameters in Figure~\ref{fig:def}. 

\subsection{\it AGN Continuum Subtraction} \label{sec:continuum}

The first step in measurement of the \oiii\ and \sii\ line profiles is
accurate subtraction of the continuum.  This is particularly important
for characterization of any broad wings on the line profiles.  In all
cases, we defined continuum regions approximately 300--600 \kms\ 
away from the lines under consideration.  For all but four of our
targets (NGC~3227, NGC~3516, NGC~4051, and NGC~4579), we found that the
faint AGN continuum was well-approximated by a constant over this narrow
wavelength range.  For these four bright galaxies, we had sufficient 
signal-to-noise ratio in the continuum to justify a power-law fit. 
The STIS slit size is sufficiently narrow that no significant stellar 
features appear to contaminate the continuum.

\subsection{\it Line Width Measurements} \label{sec:widths}

We have characterized the mean velocity structure in the \oiii\ 
$\lambda\lambda 4959,5007$ 
and \sii\ $\lambda\lambda 6717,6731$ emission lines through measurement of 
FW20, FWHM, and FW80, which are the line widths at 20\%, 50\%, and 80\% of 
the peak height, respectively.  
These parameters, as illustrated in Figure~\ref{fig:def}, yield the base 
widths, core widths, and top profile widths described in \citet{Heckman81}. 
We have linearly interpolated the flux between wavelength bins ($\sim$0.28\AA\ 
for the G750M and G430M gratings) to measure the widths.  
The value FWHM/2.354 was used for comparison with the stellar velocity 
dispersion measurements measured by \citet{Nelson95}.

Several of the AGNs have sufficiently blended \sii\ lines such that the flux 
between the 6717\AA, 6731\AA\ doublet exceeds the 20 percent peak value for 
both lines. For the most severely blended of these, the FWHM value was 
undefined as well, and for NGC~1052 and NGC~2110 the FW80 value could not be 
measured.  
We find that the \sii\ lines often become more blended with increasing aperture 
size and for several galaxies the FW20 and FWHM measurements become undefined 
above some aperture size.  In these cases we measured the width values from 
a Gaussian fit to the line profile (our Gaussian fitting scheme is described 
in \S\ref{sec:areas}).  
Table~\ref{tbl:width} lists our width measurements of \sii\ and/or \oiii\ for 
each of the sources at aperture sizes of 0.2'' and 1.0'' and indicates the 
values obtained from a Gaussian fit (with a $g$ superscript). 
Figure~\ref{fig:width} shows the value of these three quantities as a 
function of aperture size. 

\begin{figure*}
\figurenum{3}
\plotone{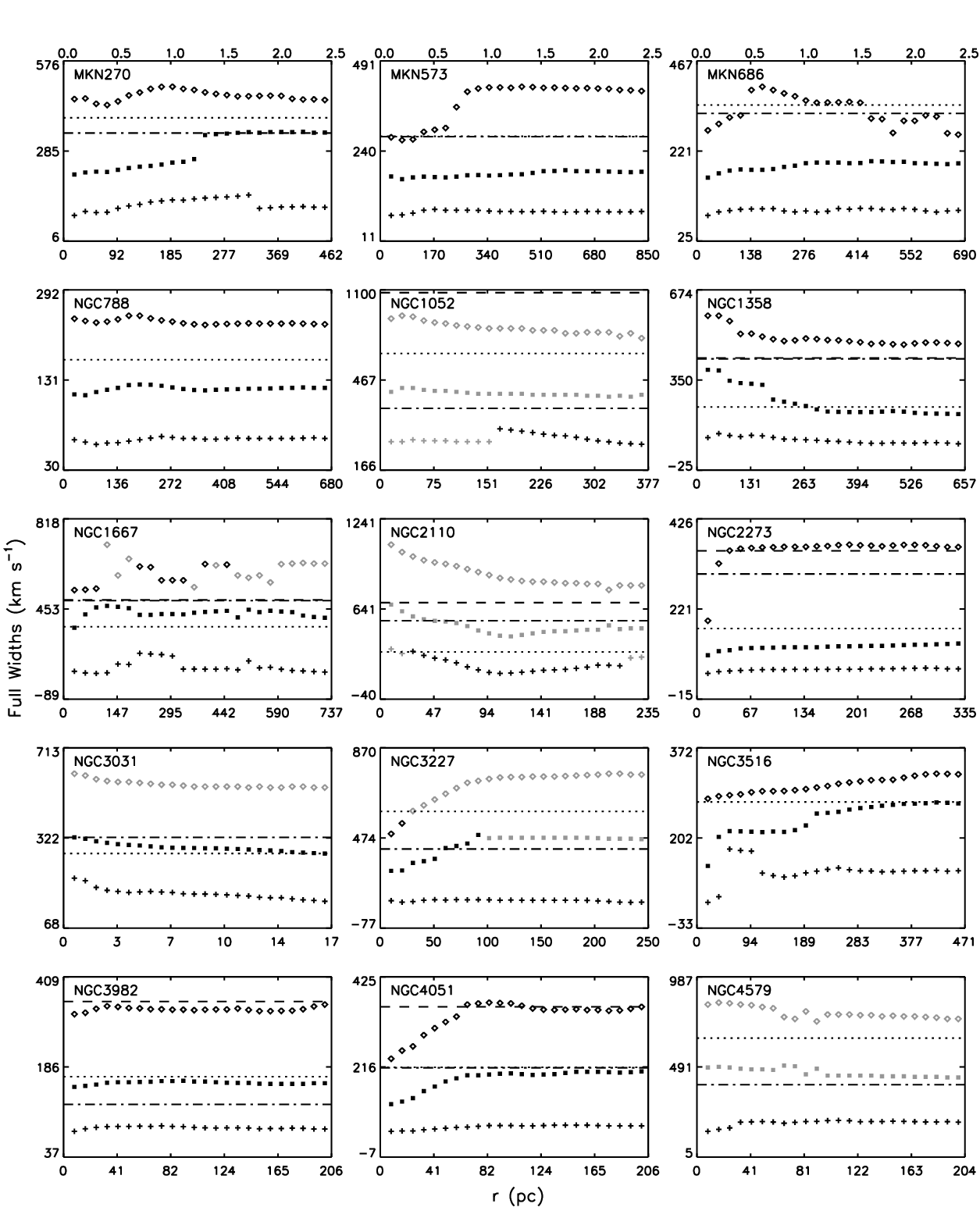} 
\caption{Line width parameters FW20 (diamonds), FWHM
(squares), and FW80 (crosses)  are plotted against aperture size in units 
of physical distance (below) and angular size (above) for each galaxy in 
our sample. The NW95 measurements for FW20 (dashed), FWHM (dotted), and 
$2.354\times\sigma_{*}$ (dash-dotted) are plotted as horizontal lines. 
Black symbols denote measured widths, while gray symbols denote values 
obtained from a Gaussian model of the line profile. 
} \label{fig:width}
\end{figure*}

\begin{figure*}
\figurenum{3}
\plotone{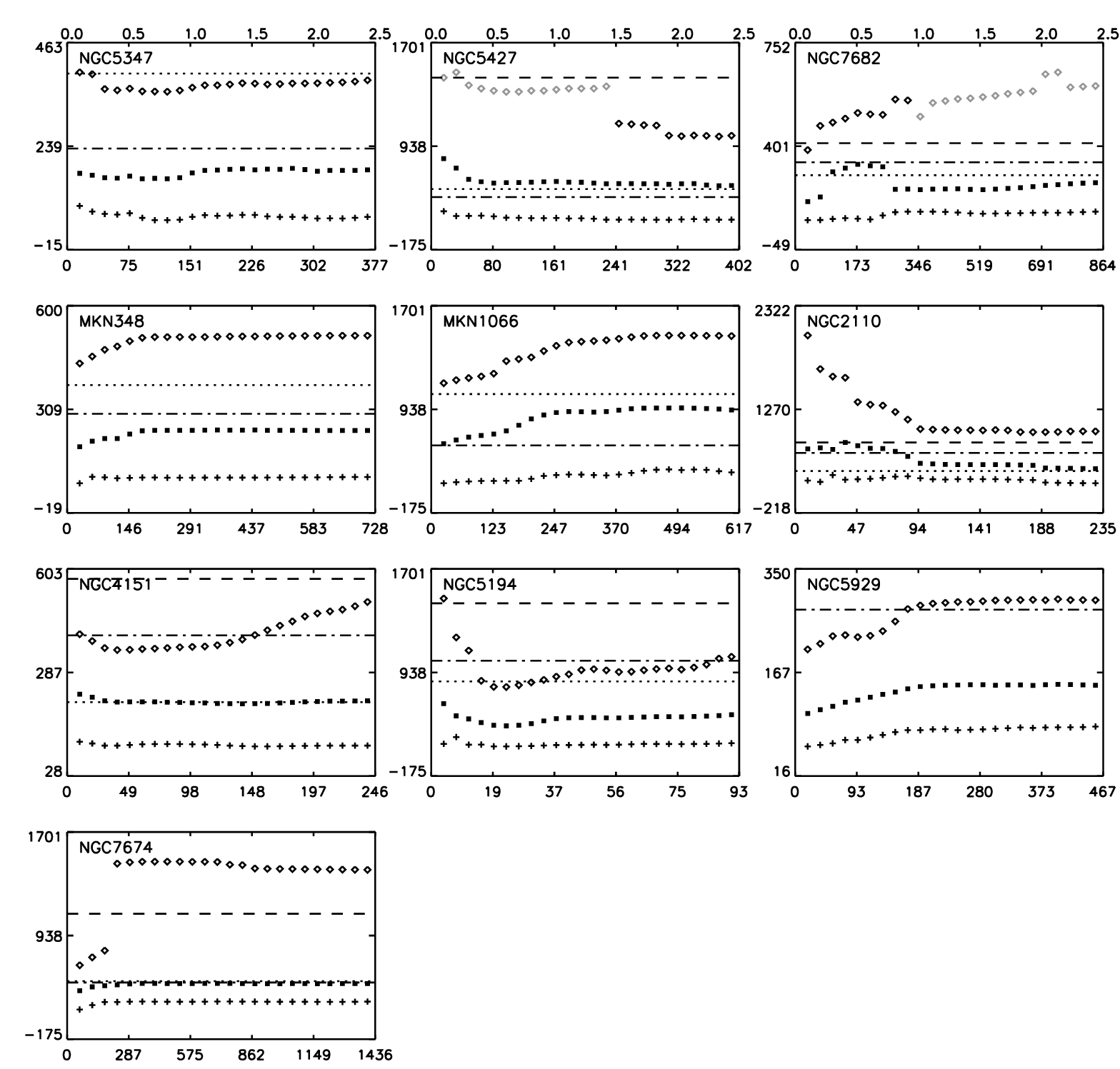} 
\caption{\it Continued}
\end{figure*}

\begin{deluxetable*}{lccccccccccccc}
\tabletypesize{\tiny} 
\tablecaption{Line Width Measurements\label{tbl:width}}
\tablenum{2}
\tablehead{
\multicolumn{1}{c}{} &
\multicolumn{4}{c}{--------- \citet{Nelson95} --------} &
\multicolumn{4}{c}{------------- Aperture $= 0.2''$ --------------} &
\multicolumn{3}{c}{-------- Aperture $= 1''$ --------} &
\multicolumn{2}{c}{} \\
\colhead{Name} & 
\colhead{$\sigma_{*}$} & 
\colhead{Line} & 
\colhead{FWHM} & 
\colhead{FW20} & 
\colhead{Line} & 
\colhead{FW80} & 
\colhead{FWHM} & 
\colhead{FW20} & 
\colhead{FW80} & 
\colhead{FWHM} & 
\colhead{FW20} & 
\colhead{$1''/0.2''$} & 
\colhead{NW95/$1''$} \\ 
\colhead{} & 
\colhead{(km/s)} & 
\colhead{} & 
\colhead{(km/s)} & 
\colhead{(km/s)} & 
\colhead{} & 
\colhead{(km/s)} & 
\colhead{(km/s)} & 
\colhead{(km/s)} & 
\colhead{(km/s)} & 
\colhead{(km/s)} & 
\colhead{(km/s)} & 
\colhead{(\%)} & 
\colhead{(\%)} \\ 
\colhead{(1)} & 
\colhead{(2)} & 
\colhead{(3)} & 
\colhead{(4)} & 
\colhead{(5)} & 
\colhead{(6)} & 
\colhead{(7)} & 
\colhead{(8)} & 
\colhead{(9)} & 
\colhead{(10)} & 
\colhead{(11)} & 
\colhead{(12)} & 
\colhead{(13)} & 
\colhead{(14)} \\
} 
\startdata
MKN 270              & 148 & [O III] & 396 & 620 & [S II]  & 100 & 223 & 457$^{g}$  & 135 & 254 & 494$^{g}$ & 14 & 56 \\
MKN 573              & 123 & [S III] & 290 & 540 & [S II]  & 81  & 176 & 280 & 91 & 186 & 420 & 5.7 & 56 \\
MKN 686              & 144 & [O III] & 359 & 609 & [S II]  & 97  & 192 & 313 & 99 & 216 & 370 & 13 & 66 \\
NGC 788              & 140 & [O III] & 190 & 295 & [S II]  & 70  & 140 & 247 & 77 & 150 & 246 & 7.1 & 27 \\
NGC 1052             & 207 & [O III] & 770 &1085 & [S II]  &314$^{g}$ & 592$^{g}$ & 966$^{g}$ & 314$^{g}$ & 612$^{g}$ & 901$^{g}$ & 3.4 & 26 \\
NGC 1358             & 173 & [O III] & 220 & 410 & [S II]  & 117 & 361 & 574 & 92 & 224 & 486 & -38 & -1.8 \\
NGC 1667	     & 173 & [O III] & 275 & 410 & [S II]  & 43  & 337 & 462 & 128 & 339 & 509 & 0.6 & -19 \\
NGC 2110  	     & 220 & [O III] & 295 & 645 & [S II] & 284$^{g}$ & 584$^{g}$ & 1005$^{g}$ &  149$^{g}$ & 426$^{g}$ & 842$^{g}$ & -27 & -31\\
NGC 2273             & 124 & [S III] & 158 & 348 & [S II]  & 52  & 102 & 317 & 58 & 112 & 358 & 9.8 & 41 \\
NGC 3031             & 167 & [O III] & 335 &- - -& [S II]  & 238 & 388 & 614$^{g}$ & 196 & 355 & 581$^{g}$ & -8.5 & -5.6 \\
NGC 3227             & 144 & [S III] & 536 &1151 & [S II]  & 59  & 225 & 473 & 71 & 395$^{g}$ & 708$^{g}$ & 76 & 36 \\
NGC 3516             & 235 & [O III] & 250 & 550 & [S II]  & 37  & 171 & 263 & 91 & 197 & 279 & 15 & 27 \\
NGC 3982             & 62  & [O III] & 203 & 358 & [S II]  & 95  & 184 & 335 & 100 & 194 & 341 & 5.4 & 4.6 \\
NGC 4051             & 88  & [S III] & 208 & 353 & [S II]  & 56  & 126 & 248 & 69 & 191 & 363 & 52 & 8.9 \\
NGC 4579             & 170 & [O III] & 653 &1278 & [S II]  & 156 & 497$^{g}$ & 847$^{g}$ & 199 &  456$^{g}$ & 799$^{g}$ & -8.2 & 43\\
NGC 5347             & 93  & [O III] & 392 & 677 & [S II]  & 73  & 159 & 393 & 61 & 165 & 360 & 3.8 & 140 \\
NGC 5427             & 74  & [O III] & 264 & 620 & [S II]  & 103 & 282 & 641$^{g}$ & 96 & 232 & 576$^{g}$ & -18 & 14 \\
NGC 7682             & 123 & [S III] & 239 & 363 & [S II]  & 66  & 155 & 431 & 98 & 183 & 466$^{g}$ & 18 & 31 \\
 &  &  &  &  &  &  &  &  &  &  &  &  &  \\
MKN 348              & 118 & [O III] & 363 & 660 & [O III] & 89  & 195 & 448 & 86 & 228 & 506 & 17 & 59 \\
MKN 1066             & 105 & [O III] & 417 & 714 & [O III] & 124 & 265 & 463 & 148 & 356 & 578 & 34 & 17 \\
NGC 2110 	     & 220 & [O III] & 295 & 645 & [O III] & 161 & 575 & 1546 & 209 & 394 & 855 & -31 & -25 \\
NGC 4151             & 178 & [S III] & 233 & 575 & [O III] & 118 & 246 & 403 & 116 & 231 & 387 & -6.1 & 0.9 \\
NGC 5194             & 102 & [S III] & 195 & 364 & [O III] & 75  & 120 & 291 & 57 & 115 & 205 & -4.2 & 70 \\
NGC 5929             & 121 & [O III] & 405 & 576 & [O III] & 66  & 123 & 230 & 90 & 160 & 292 & 30 & 153 \\
NGC 7674             & 144 & [O III] & 350 & 960 & [O III] & 133 & 298 & 566 & 164 & 328 & 1430 & 10.1 & 6.7 \\
\enddata
\tablecomments{Col. (1): Galaxy name. Col. (2): Stellar velocity 
dispersion measurement from NW95. Col. (3) Line used for width measurements 
in NW95. Cols. (4)--(5): FWHM and FW20 values from NW95. Col. (6): Line used 
for our width measurements. Cols. (7)--(9): Our FW80, FWHM, and FW20 
measurements in an $0.2''$ aperture (the $g$ superscript denotes measurements 
obtained from a Gaussian model of the line profile). 
Cols. (10)--(12): FW80, FWHM, and FW20 values measured in a $1''$ aperture. 
Col. (13): The percent change in FWHM between the $0.''2$ to $1''$ aperture 
sizes. Col. (14): The percent change between our $1''$ FWHM measurement and 
the NW95 value.}
\end{deluxetable*}

These width measurements are the observed widths and have not 
been corrected for the instrumental resolution. As noted in 
section~\ref{sec:sample}, the FWHM velocity resolution of the G430M and 
G750M gratings are $\sim 25$ and 40 \kms, respectively, for an unresolved 
point source and a factor of up to $\sim 2.5$ times higher for a source 
that fills the slit. The most commonly used method to correct for 
broadening due to the instrumental resolution is to subtract 
the width of the instrumental resolution in quadrature. 
This correction is can not be performed with great accuracy for these 
observations because there is a substantial difference in resolution between 
unresolved and resolved emission and both sources of emission likely 
contribute to the observed emission line profile. However, we can estimate 
the magnitude of this 
effect for the narrowest lines in our sample, which have FWHM $\sim 120$ 
\kms\ (see Table~\ref{tbl:width}), and represent the most affected 
measurements. If this emission were completely unresolved the correction 
would be less than 5\% and if it uniformly filled the slit it would be less 
than 30\%. Because the two-dimensional spectra of these sources all indicate 
that the emission-line gas is centrally peaked, we estimate that the majority 
of the emission in the smallest aperture is still unresolved. We therefore 
conclude that instrumental resolution makes at most a small contribution 
to our reported line widths. For the majority of our sample, the line widths 
are sufficiently broad that this correction is less than 10\% even with the 
most extreme assumption that the emission uniformly fills the aperture. We 
also note that the instrumental resolution correction is likely to be more 
important for the line core (FW80) than at smaller fractions of the peak 
height \citep[e.g.][]{Whittle85a}. 

In their study of 1749 AGNs spectra from the SDSS, \citet{Greene05} 
found that the moments of Gaussian fits to the profiles provide more 
robust widths than direct measurements from the spectra. However, the SDSS 
data are typically lower signal-to-noise ratio (SNR) and width 
measurements can be more uncertain in the low SNR regime. The STIS spectra 
used for our sample generally have sufficient SNR and resolution that
we can confidently measure emission line widths without assuming any
model for the line shape.  Also, because the majority of the line
profiles have a blue or redshifted wing and/or a broad central
component, measuring widths directly from the data allows us to avoid
approximating single values of the moments from a combination
of multiple Gaussian components. 

The dominant uncertainty in direct measurement of the line widths 
is the continuum level because these lines have high signal-to-noise 
ratio. As noted above in Section~\ref{sec:continuum}, we fit the continuum 
by either a constant or a power law.
The line width measurement uncertainty depends on both the 
uncertainty in the continuum level and the line profile shape, as for example 
a given continuum uncertainty will produce a smaller width uncertainty in 
a broad line than a narrow line. We estimate that the uncertainties in 
our direct width measurements are less than 5\%. The width measurements 
that required Gaussian fits have a formal uncertainty of approximately 
5\% \citep{Greene05}, although this is only a true estimate if the 
lines can be correctly represented by Gaussians. 
Those cases where the fitting routine varied between direct width measurements 
and Gaussians (the switch between black and gray symbols in 
Figure~\ref{fig:width}) indicate that the uncertainties are at most 10\%.
The lack of substantial stellar continuum emission in the extremely 
narrow STIS slit removes a potentially significant source of the 
uncertainty in width measurements based on ground-based observations, 
which may be particularly relevant for fainter emission lines such as \sii. 

\subsection{\it Area and Asymmetry Measurements} \label{sec:areas}

\citet{Whittle85a} advocates the use of area measurements to define line
width parameters, rather than simple cuts at varying heights, because they 
have an integral nature and are thus smoothly defined and less sensitive to 
the presence of noise or the effects of instrument resolution. 
His definitions of the interpercentile velocity
widths (IPV10 and IPV20) and asymmetries (A10 and A20) are also
illustrated in Figure~2. The median is the wavelength that denotes
the center of area for the profile, and the lengths a, b, c, and d
represent the separation between the median and the profile's 10\%, 90\%, 
20\%, and 80\% area values, respectively.  The IPV10 parameter characterizes 
the base and wings of the profile, much as did the FW20 width suggested by 
\citet{Heckman81}, while the IPV20 parameter and higher percentage areas 
characterize the line core. The A10 and A20 values serve to clearly quantify 
the profile's red or blue asymmetry. 

We have chosen to measure our area parameters from Gaussian fits to
the line profiles, rather than directly from the data as discussed in
\citet{Whittle85a}. This choice was driven by the \sii\ lines, which are mildly 
to severely blended in all of our sources. 
Because these lines are blended, we could not 
measure the interpercentile markers in the same manner as \citet{Whittle85a} 
for the more isolated \oiii\ $\lambda 5007$ line. 
We experimented with taking the 
10 and 20 percentile markers of the total \sii\ doublet area from the 
blue wing of the 6717\AA\ line, and the 80 and 90 percentile markers from the 
red wing of the 6731\AA\ line, and then scaling the widths a, b, c and d by the
relative widths of the \sii\ lines, but the potentially variable line ratio 
of the \sii\ doublet made this approach uncertain. In the end, we found that 
the Gaussian profiles provided more robust measurements of the area 
parameters, as we were able to isolate the separate contributions to the 
line profile from each of the \sii\ lines.  Also, the Gaussian fits allowed 
us to avoid the subtleties in treating fractional wavelength increments 
discussed by \citet{Whittle85a}.

Very few of these galaxies are well fit by a single Gaussian.  
The \oiii\ line often has an asymmetric blue wing in addition to a broad 
central component, and so we allow for up to three Gaussian components in our 
fitting routines. With the exception of the sources with bright AGN continua, 
we fit a constant continuum term.  We fit the \oiii\ $\lambda\lambda 4959$ 
and 5007 
lines simultaneously with their wavelength separation and relative strengths 
fixed to the theoretical ratio of 3:1 determined by atomic physics.  
For the \sii\ doublet we also fit the two 
lines simultaneously and with a fixed separation, although we allow their
relative line strengths to vary because the \sii\ $\lambda$6717/$\lambda$6731
ratio is sensitive to electron density. 
In principle, the line widths of the 6717\AA\ and 6731\AA\ lines may be 
different due to stratification in the NLR, so we initially allowed the 
widths of the two lines to vary in our fitting routine. However, we 
found that the widths of the two lines differed by less than 3\%, which is 
within the error of the Gaussian fit parameters described in the previous 
section. We therefore fixed the widths of these two lines in the 
fitting routine. 

The fitting routine employed here was adopted from \citet{Greene05}. 
The first step is to fit a single Gaussian to the profile, and then we allow 
a second central component centered between -5\AA\ and +5\AA\ of the line
peak (although occasionally the component was best fit outside these limits). 
We keep this second component if the $\chi^{2}$ value for the fit 
improves by at least 20\%. For \oiii\ we experimented with a third, blue 
component that was generally centered between -20\AA\ and 0\AA\ of the central 
component. Again we kept this component when $\chi^{2}$ improved by at least 
20\%. For the \sii\ lines we only allowed for one additional component 
between -10\AA\ and +10\AA\ of the line peak. 
The best fits from our Gaussian routine for each forbidden line profile are 
shown in Figure~\ref{fig:spectra} and the fit residuals are plotted below 
the spectrum in each panel. 
The IPV at $0.''2$ and $1''$ are provided in Table~\ref{tbl:ipv} and plotted 
as a function of aperture size in Figure~\ref{fig:ipv}, while the 
asymmetry measurements are listed in Table~\ref{tbl:asym} and shown in 
Figure~\ref{fig:asym}. Unlike the width measurements, the IPV area-defined 
parameters are relatively insensitive to instrumental resolution 
\citep{Whittle85a}.  

\begin{figure*}
\figurenum{4}
\plotone{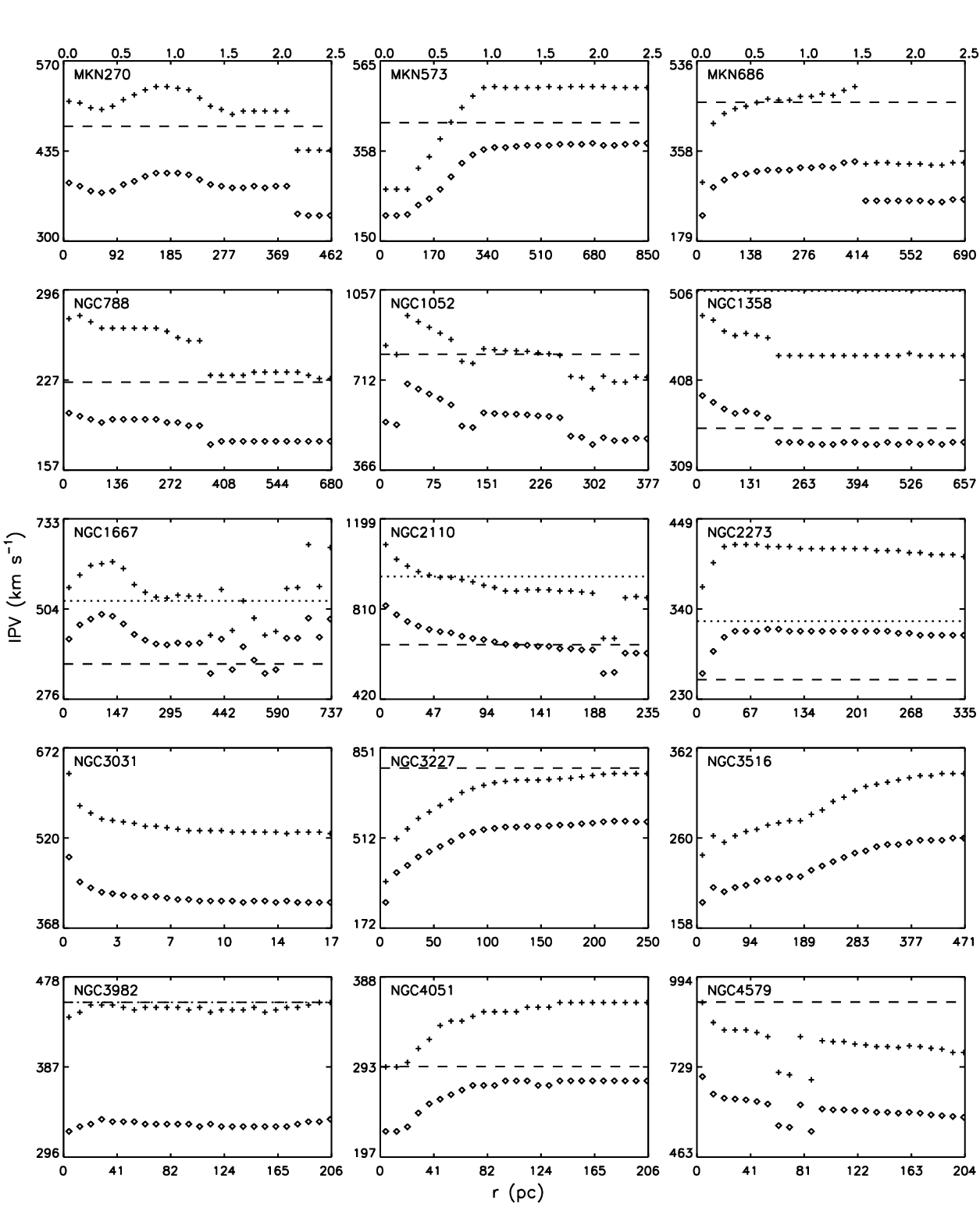}
\caption{\footnotesize IPV10 (crosses) and IPV20 (diamonds) measurements 
obtained from the Gaussian fits to the line profiles plotted against aperture 
size as in Figure~\ref{fig:width}. The \citet{Nelson95} measurements for
IPV10 (dotted) and IPV20 (dashed) are plotted as horizontal lines. NGC~7682 
is not shown because the red wing of the \sii\ line did not fall on the 
STIS CCD. 
} \label{fig:ipv}
\end{figure*}

\begin{figure*}
\figurenum{4}
\plotone{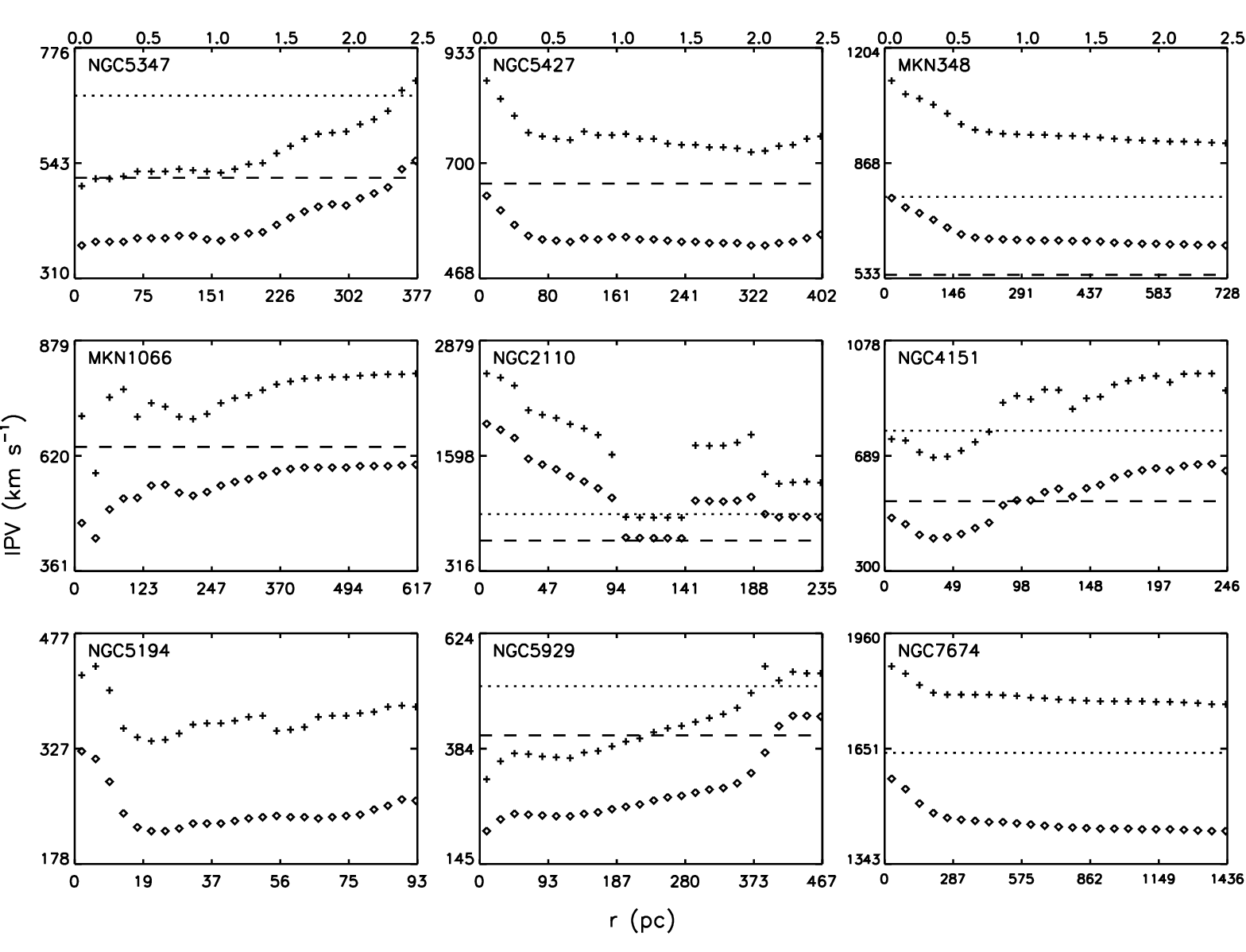} 
\caption{\it Continued}
\end{figure*}

\begin{deluxetable*}{lcccccccccc}
\tabletypesize{\scriptsize}
\tablecaption{IPV Width Measurements\label{tbl:ipv}}
\tablenum{3}
\tablehead{
\multicolumn{1}{c}{} &
\multicolumn{3}{c}{----\citet{Nelson95}----} &
\multicolumn{1}{c}{} &
\multicolumn{2}{c}{----Aperture $= 0.2''$----} &
\multicolumn{2}{c}{----Aperture $= 1''$----} &
\multicolumn{2}{c}{----IPV20----} \\
\colhead{Name} & 
\colhead{Line} & 
\colhead{IPV20} & 
\colhead{IPV10} & 
\colhead{Line} & 
\colhead{IPV20} & 
\colhead{IPV10} & 
\colhead{IPV20} & 
\colhead{IPV10} & 
\colhead{$1''/0.2''$} & 
\colhead{NW95/1$''$} \\ 
\colhead{} & 
\colhead{} & 
\colhead{(\kms)} & 
\colhead{(\kms)} & 
\colhead{} & 
\colhead{(\kms)} & 
\colhead{(\kms)} & 
\colhead{(\kms)} & 
\colhead{(\kms)} & 
\colhead{(\%)} & 
\colhead{(\%)} \\
\colhead{(1)} & 
\colhead{(2)} & 
\colhead{(3)} & 
\colhead{(4)} & 
\colhead{(5)} & 
\colhead{(6)} & 
\colhead{(7)} & 
\colhead{(8)} & 
\colhead{(9)} & 
\colhead{(10)} & 
\colhead{(11)} \\
} 
\startdata
MKN 270              & [O III] & 472 & 637  & [S II]  & 238 & 371 & 250 & 393 & 5.0 & 89  \\
MKN 573              & [S III] & 423 & 572  & [S II]  & 132 & 211 & 196 & 343 & 48  & 120 \\
MKN 686              & [O III] & 454 & 593  & [S II]  & 153 & 273 & 192 & 333 & 25  & 140 \\
NGC 788              & [O III] & 225 & 305  & [S II]  & 104 & 182 & 117 & 191 & 13  & 92  \\
NGC 1052             & [O III] & 810 & 1075 & [S II]  & 700 & 961 & 584 & 829 & -17 & 39  \\
NGC 1358             & [O III] & 355 & 505  & [S II]  & 339 & 477 & 254 & 383 & -25 & 40  \\
NGC 1667	     & [O III] & 365 & 525  & [S II]  & 549 & 714 & 423 & 549 & -23 & -14 \\
NGC 2110             & [S III] & 655 & 950  & [S II]  & 766 &1010 & 673 & 907 & -12 & -3  \\
NGC 2273             & [S III] & 254 & 325  & [S II]  & 160 & 288 & 189 & 303 & 18  & 34  \\
NGC 3031             & [O III] &- - -&- - - & [S II]  & 441 & 567 & 416 & 537 & -5.7& - - - \\
NGC 3227             & [S III] & 775 & 994  & [S II]  & 394 & 526 & 546 & 715 & 39  & 42 \\
NGC 3516             & [O III] & 490 & 665  & [S II]  & 202 & 258 & 221 & 282 & 9.4 & 120 \\
NGC 3982             & [O III] & 452 & 452  & [S II]  & 189 & 324 & 198 & 335 & 4.8 & 130 \\
NGC 4051             & [S III] & 293 & 405  & [S II]  & 224 & 293 & 273 & 351 & 22  & 7.3 \\
NGC 4579             & [O III] & 920 & 1143 & [S II]  & 586 & 750 & 547 & 704 & -6.7& 68  \\
NGC 5347             & [O III] & 513 & 679  & [S II]  & 236 & 392 & 233 & 395 & -1.3& 120 \\
NGC 5427             & [O III] & 659 & 956  & [S II]  & 299 & 554 & 278 & 518 & -7.0& 140 \\
NGC 7682             & [S III] & 266 & 330  & [S II]  & 774 &1102 & 784 &1136 & 1.3 & -66 \\
 &  &  &  &  &  &  &  &  &  &  \\
MKN 348              & [O III] & 543 & 770  & [O III] & 669 & 991 & 625 & 920 & -6.6& -13 \\
MKN 1066             & [O III] & 640 & 902  & [O III] & 445 & 596 & 545 & 725 & 22  & 17  \\
NGC 2110             & [O III] & 655 & 950  & [O III] &1670 &2199 &1105 &1640 & -33 & -41  \\
NGC 4151             & [S III] & 536 & 774  & [O III] & 441 & 721 & 533 & 874 & 21  & 0.6 \\
NGC 5194             & [S III] &- - -&- - - & [O III] & 299 & 422 & 232 & 362 & -22 & - - - \\
NGC 5929             & [O III] & 412 & 514  & [O III] & 254 & 409 & 261 & 397 & 2.8 &  58 \\
NGC 7674             & [O III] & 1255 & 1640 &[O III] &1524 &1836 &1444 &1784 & -5.2& -13 \\
\enddata
\tablecomments{Col. (1): Galaxy name. Col. (2) Line used for
width measurements in NW95. Cols. (3)--(4): IPV20 and IPV10 values
given by NW95.  Col. (5): Line used for our width
measurements. Cols. (6)--(7): IPV20 and IPV10 values for our sample
with a 0.''2 aperture. Cols. (8)--(9): IPV20 and IPV10 values for our
sample with a 1'' aperture. Col. (10): The percent change in IPV20
from 0.''2 to 1'' aperture sizes. Col. (11): The percent difference
between our IPV20 measurement within 1'' and the NW95 IPV20 value.
}
\end{deluxetable*}

\begin{deluxetable*}{lcccccc}
\tablecaption{Asymmetry Measurements\label{tbl:asym}}
\tablenum{4}
\tabletypesize{\scriptsize}
\tablehead{
\colhead{Name} &
\colhead{Line} &
\colhead{A10 (0."2)} &
\colhead{A20 (0.''2)} &
\colhead{A10 (1'')} &
\colhead{A20 (1")} &
\colhead{Blue Area (1"-0."2)} \\
\colhead{} &
\colhead{} &
\colhead{(\kms)} &
\colhead{(\kms)} &
\colhead{(\kms)} &
\colhead{(\kms)} &
\colhead{(\%)} \\
\colhead{(1)} &
\colhead{(2)} &
\colhead{(3)} &
\colhead{(4)} &
\colhead{(5)} &
\colhead{(6)} &
\colhead{(7)} \\
}
\startdata
MKN 270    & [S II] & -0.17 & -0.20 & -0.12 & -0.14 &  0.43  \\
MKN 573    & [S II] & 0.00 & -0.01 & -0.21 & -0.22 &  0.24  \\
MKN 686    & [S II] & -0.3 &-0.03 & 0.00 & 0.00 &  - - -   \\
NGC 788    & [S II] & 0.13 & 0.11 & 0.03 & 0.03 & 0.76  \\
NGC 1052   & [S II] & -0.09 & -0.09 & -0.05 & -0.06 & 0.35  \\
NGC 1358   & [S II] & 0.01 & 0.03 & 0.01 & 0.01 & - - -   \\
NGC 1667   & [S II] & -0.23 & -0.21 & 0.03 & 0.04 & 0.46   \\
NGC 2110   & [S II] & -0.09 & -0.10 & 0.00 & 0.00  & 0.48  \\
NGC 2273   & [S II] & 0.34 & 0.42 & 0.34 & 0.39 &  0.76  \\
NGC 3031   & [S II] & 0.00 & 0.01 & 0.00 & -0.01 & 0.77 \\
NGC 3227   & [S II] & 0.09 & 0.12 & 0.10 & 0.13 &  1.00 \\
NGC 3516   & [S II] & 0.00 & 0.01 & 0.00 & 0.01 & - - -  \\
NGC 3982   & [S II] & 0.26 & 0.25 & 0.24 & 0.22 &  - - -  \\
NGC 4051   & [S II] & 0.29 & 0.31 & 0.25 &0.25 &  0.82 \\
NGC 4579   & [S II] & -0.06 & -0.07 & -0.08 & -0.09 &  0.59  \\
NGC 5347   & [S II] & 0.35 & 0.37 & 0.32 & 0.34 &  0.56 \\
NGC 5427   & [S II] & -0.02 & -0.02 & -0.02 & -0.02 & - - -\\ 
NGC 7682   & [S II] & - - - & - - -  & - - - & - - - & 0.52 \\
            &  &  &  &  \\
MKN 348    & [O III] & 0.36 & 0.38 & 0.28 & 0.29 & 0.56  \\
MKN 1066   & [O III] & 0.41 & 0.33 & 0.31 & 0.29 &  0.46  \\
NGC 2110   & [O III] & 0.15 & 0.12 & 0.11 & 0.14 &  0.78  \\ 
NGC 4151   & [O III] & -0.02 & -0.02 & 0.14 & 0.12 & 0.42  \\
NGC 5194   & [O III] & -0.13 & -0.18 & 0.03 & 0.06 & 0.52  \\
NGC 5929   & [O III] & -0.13 & -0.17 & -0.08 & -0.06 &  0.70 \\
NGC 7674   & [O III] & 0.46 & 0.44 & 0.45 & 0.45 & 0.75 \\
\enddata
\tablecomments{Col. (1): Galaxy name.  Col. (2): Emission line used
in STIS data analysis.  Col. (3) -- (4): A20 and A10 asymmetry measurement with 
an $0.2''$ aperture.  Col. (5) -- (6): A20 and A10 measurement within a
$1''$ aperture. Col. (7): Fraction of the 1''-0.''2 profile area that lies 
blueward of the 1'' line centroid.}
\end{deluxetable*}

\begin{figure*}
\figurenum{5}
\plotone{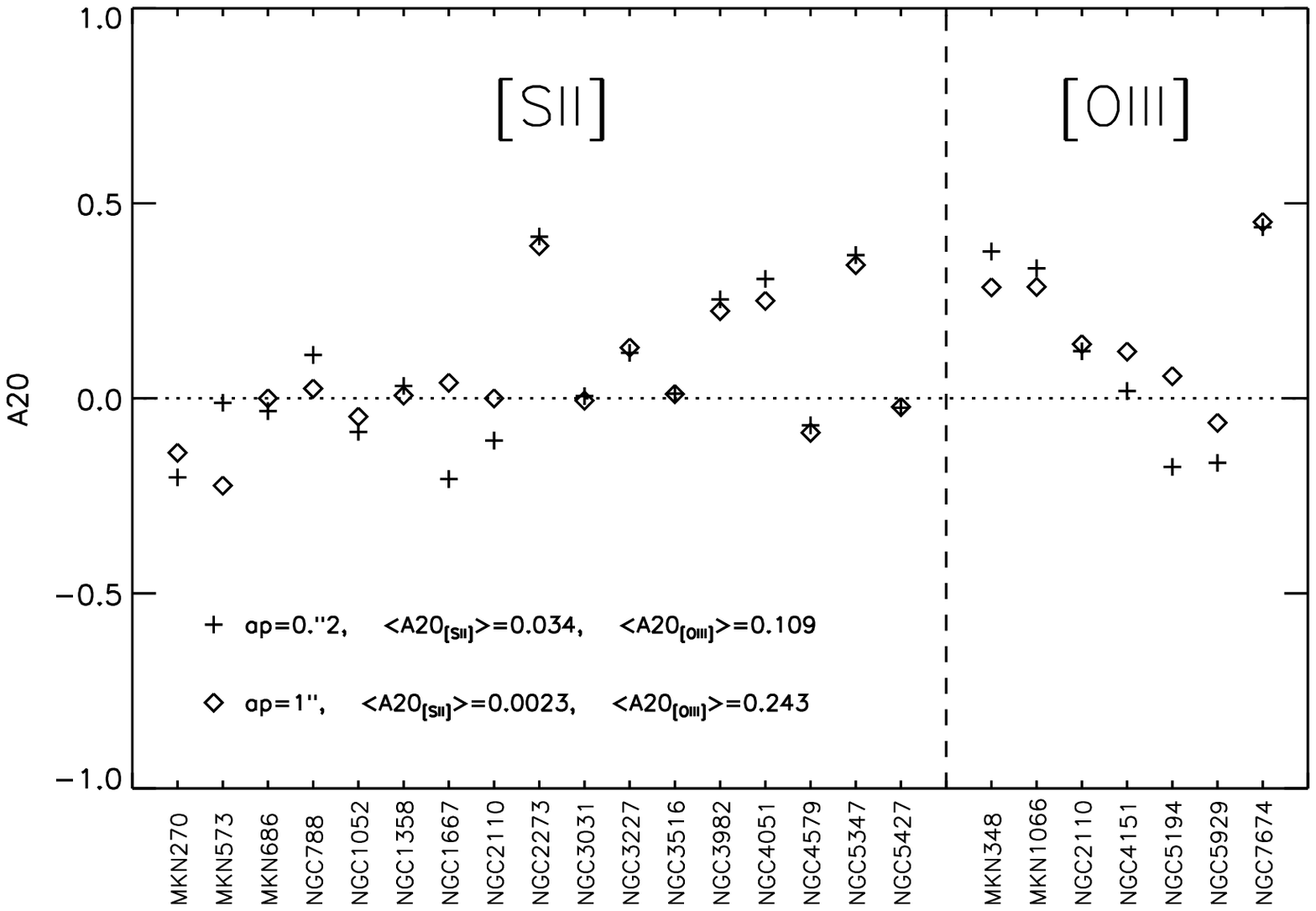} 
\caption{A20 measurements for the \sii\ and \oiii\ emission lines in our
sample at aperture sizes of $0.2''$ (crosses) and $1''$ (diamonds). 
Positive A20 values indicate blue asymmetries, negative A20 values indicate 
red asymmetries, and near-zero values indicate symmetric profiles.
The average A20 values for \sii\ and \oiii\ at each aperture are
shown. \label{fig:asym} }
\end{figure*}

\section{Results}  \label{sec:res}

All but three of the 24 galaxies in the sample have measurable emission 
outside of a central, $0.2''$ aperture, where $0.2''$ corresponds to 8 -- 115 
parsecs for this sample (excluding the very nearby NGC~3031). The three 
galaxies that lack detectable emission outside of $0.2''$ (MKN~686, NGC~3516, 
and NGC~5427) were observed with G750M and not G430M. The brighter 
\oiii\ line is always detectable outside of the central $0.2''$. 
The presence of significant emission outside of $0.2''$ is 
illustrated in the rightmost columns of Figure~\ref{fig:spectra}, which 
displays the difference between the sum of a $1''$ and $0.2''$ aperture 
centered on the nucleus. In addition to the three galaxies that do not 
exhibit detectable emission outside of $0.2''$, nine additional galaxies 
with \sii\ observations have weak emission on larger scales and six have 
substantial emission. 
However, even emission lines that are quite weak outside of $0.2''$ can 
contribute to the line profiles in larger, integrated apertures. 

\subsection{Line Widths vs.\ Aperture Size} 

The line width and IPV values as a function of aperture size are shown in 
Figures~\ref{fig:width} and \ref{fig:ipv}, respectively. 
Figure~\ref{fig:width} plots the radial dependence of FW20, FWHM, and FW80 
for each galaxy, while Figure~\ref{fig:ipv} shows IPV10 and IPV20. 
The figures also include the ground-based measurements from NW95 as 
horizontal lines, unless the values measured in the ground-based aperture 
fall outside the range of the vertical axes. 
Our measurements in $0.2''$ and $1''$ apertures, along with the NW95 data, 
are listed in Tables~\ref{tbl:width} and \ref{tbl:ipv}. For nearly all of 
the galaxies there are substantial differences between the values in the 
nuclear region and the maximum STIS aperture size, as well as between the 
maximum STIS aperture size and the NW95 measurements. 
These differences between the STIS apertures and the ground-based measurements 
could be due to uncertain resolution corrections in the narrowest lines, 
although galaxies with narrow emission lines are not systematically more 
different than galaxies with well-resolved lines. 
An alternate explanation is that these measurements fall below those from 
NW95 because the STIS slit only subtends a fraction of the NLR. 
We will discuss this point further in Section~\ref{sec:dis}. 

Figure~\ref{fig:width} shows that the line profile widths almost always 
increase or remain approximately constant. The FWHM for ten galaxies 
increases by greater than 10\%, while for an additional 11 the FWHM 
changes by less than 10\%. Only three galaxies decrease by greater than 
10\% from the $0.2''$ to $1''$ aperture (NGC~1358, NGC~5427, and NGC~2110 
in both emission lines).  
Two galaxies (NGC~1358 \sii, NGC~2110 \oiii) decrease by more than $30$\%  
and are interesting cases because a decrease requires both significant 
emission at larger spatial scales and a velocity width that is substantially 
narrower at larger scales than in the nuclear region. Comparison of the 
$0.2''$ and $1''$ apertures clearly indicates that the profiles are 
substantially broader in the central region. 

IPV variations with aperture size are particularly sensitive to changes in the 
base and wings of the lines, which may probe the acceleration or deceleration 
of winds. 
For example, an increase corresponds to more high-velocity emission-line 
gas outside of the nuclear region than in the nucleus. 
While eleven of these galaxies have IPV20 variations of less than 10\% between 
the $0.2''$ and $1''$ STIS apertures, several exhibit substantial variations. 
IPV20 increases by over 30\% from $0.2''$ to $1''$ in both MKN~573 and 
NGC~3227, while it decreases by over 30\% for NGC~2110 \oiii\ (and nearly 
this amount in \sii). 
In contrast, our FW20 measurements for these galaxies only increase moderately 
with aperture size, if at all, which implies that the IPV width measurements 
are indeed more sensitive to behavior in the wings, as was suggested by 
\citet{Whittle85a}. The substantial IPV decreases with radius in NGC~1358 
and NGC~2110 provide good evidence for radiation-driven winds that decelerate 
at larger scales, as has already been noted for NGC~4151 by \citet{Nelson00b}.

We note that several IPV profiles in Figure~\ref{fig:ipv} exhibit pronounced 
jumps in IPV value with radius between neighboring data values. 
These jumps are either due to instances where our Gaussian fitting routine 
switched between a single-component fit and a multiple-component 
fit (or vice versa), or the presence of knots of line-emitting gas outside 
of the nuclear region. 

\subsection{Asymmetries}  \label{sec:asym} 

Figures~\ref{fig:width} and \ref{fig:ipv} described above indicate that 
markedly different velocity components contribute to the emission-line 
profiles in the nucleus and at larger scales in approximately half of the 
sample. In addition to this information on the widths of these velocity 
components, the presence or absence of asymmetries can provide information on 
the origin of the line-emitting material along the line of sight, particularly 
in the presence of significant gas and dust in the NLR. 

As noted previously, the \oiii\ line is quite often reported to have 
significant, typically blue, asymmetries, while asymmetries are rarely 
observed in lower-excitation lines \citep{Heckman81,deRobertis84,Whittle85b}. 
Our measurements of the \citet{Whittle85b} asymmetry parameters listed in 
Table~\ref{tbl:asym} confirm that asymmetries are more common in \oiii\ 
and these values are illlustrated in Figure~\ref{fig:asym}. 
Within the $0.2''$ aperture we find that 6/7 galaxies with \oiii\ observations 
have $|A_{20}| > 0.1$, while 9/18 galaxies with \sii\ observations have such 
asymmetries. While we observe asymmetries less frequently in \sii\ than 
\oiii, we still measure asymmetries in a substantially larger fraction 
of galaxies than typically observed in ground-based observations and we 
discuss this point further below. 
Two particularly extreme cases of asymmetries are NGC~1667 \sii\ and 
NGC~5194 \oiii.
The asymmetries are less pronounced in the larger $1''$ aperture, where 
only 7/18 galaxies in the \sii\ sample and 5/7 in the \oiii\ sample have 
such significant asymmetries. 

Several of these galaxies exhibit rather unusual asymmetries. 
While most have blue asymmetries, several have significant red asymmetries. 
One peculiar case is MKN~573, which is more asymmetric on larger scales 
{\it and} this larger-scale asymmetry is red: $A_{20} = -0.22$ in the $1''$ 
aperture. 
Inspection of \hst\ images of MKN~573 \citep{Martini03} reveals 
that the strong red asymmetry outside of the nucleus is due to the 
chance intersection of the STIS slit with an individual NLR cloud. 
MKN~270 \sii, NGC~1667 \sii, and NGC~5194 \oiii\ also exhibit significant red 
asymmetries in the $0.2''$ aperture spectra, although are less asymmetric on 
larger scales. The red-asymmetric component in these cases is broader than 
the core. 

Blue asymmetries in \oiii\ were suggested by \citet{Heckman81} to be due to 
a combination of radial outflow from the nuclear region and gas and dust 
that obscures the redshifted emission from the far side of the galaxy. 
The blue asymmetries confined to the nuclear region in NGC~4051, NGC~5347,
and MKN~348 agree well with this interpretation. NGC~7682 and MKN~1066 
have pronounced asymmetries on larger scales as well (see 
Figure~\ref{fig:spectra}). For these two galaxies $1''$ corresponds to 
several hundred parsecs and would require a large obscuring medium. 

We have also investigated the scale-dependence of asymmetries with the 
spectra shown in the rightmost panels of Figure~\ref{fig:spectra}. 
These spectra are the difference between the spectrum of each galaxy summed 
with a $1''$ aperture and a $0.2''$ aperture and therefore effectively 
isolate the emission-line component due to material outside of the nuclear 
region. The two vertical lines in each panel for a given galaxy correspond to 
the line peak of the emission lines in the $1''$ aperture. The line profiles  
in the rightmost panels relative to these vertical lines therefore illustrate 
the extent to which emitting material outside of $0.2''$ contributes to any 
asymmetries in the line profiles. To characterize the line asymmetry outside 
of the nuclear, $0.2''$ aperture, we calculated the fraction of the total 
$1'' - 0.2''$ line flux on the blue side of the line centroid in the $1''$ 
aperture. A fraction greater than 0.5 corresponds to a blue asymmetry, while
less than 0.5 corresponds to a red asymmetry. 
These values are listed in the last column of Table~\ref{tbl:asym}. 
Eight galaxies have strong blue asymmetries outside of $0.2''$ 
(fraction $> 0.6$), two have strong red asymmetries (fraction $<0.4$), and 
the remaining 14 are either relatively symmetric (ten) or have little flux 
outside of $0.2''$ (five) in the STIS aperture. The significant asymmetry in 
NGC~2110 is only seen in \oiii. 

\section{Discussion}  \label{sec:dis}

The changes in the NLR velocity field in the STIS aperture have revealed 
pronounced differences between the unresolved, nuclear kinematics and the 
NLR on larger scales, yet these larger-scale kinematics still subtend only 
a fraction of the NLR for many galaxies because of the narrow STIS slit. 
Even the line characteristics measured from the integrated STIS slit may 
therefore differ from ground-based flux measurements. 
We estimated the fraction of the NLR observed with these observations 
through a comparison of the total flux in the STIS aperture and the value 
reported in the ground-based measurements from \citet{Whittle92a}. 
This comparison demonstrated that the STIS aperture includes between 25\% and 
90\% of the \oiii\ flux measured by \citet{Whittle92a} for the seven galaxies 
with STIS \oiii\ measurements. 
The velocity field sampled by the STIS apertures also may depend on the 
orientation of the spectroscopic slit relative to the major axis 
of the host galaxies, as there is evidence that rotation is partly 
responsible for the NLR widths \citep{Whittle92b}, or the orientation relative 
to the NLR and radio jet axis, which are known to be unrelated to the 
host galaxy semimajor axis \citep{Ulvestad84,Schmitt03}. 
However, the orientation for most of these observations were not specified 
in order to reduce scheduling constraints. 

\subsection{Comparison with Ground-based Measurements} 

\begin{figure*}
\figurenum{6}
\plotone{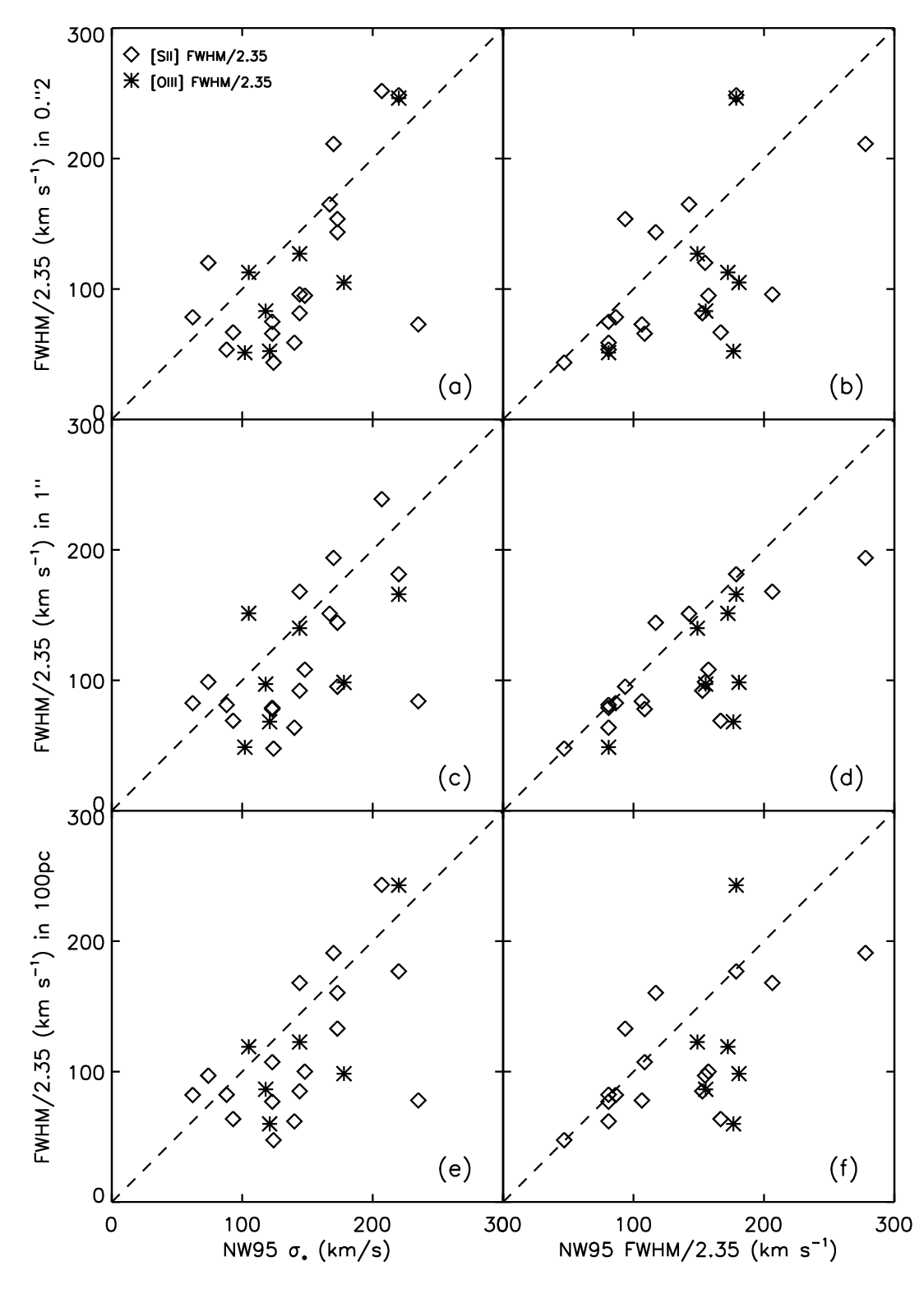} 
\caption{Our measurements of FWHM/2.354 plotted against 
NW95 $\sigma_*$ (left panels) and NW95 FWHM/2.354 (right panels). 
The top panels (a \& b) show our \sii\ (diamond) and \oiii\ (star) FWHM 
measurements in an $0.2''$ aperture, the middle two panels show our 
measurements in a $1''$ aperture, and the bottom two panels show our 
measurements in a fixed physical aperture size of 100pc. 
The dashed line in each frame represents a 1:1 correspondence between the
quantities plotted. 
\label{fig:bigcomp} }
\end{figure*}

One striking characteristic of Figures~\ref{fig:width} and \ref{fig:ipv} is 
how poorly the line profile measurements agree with the ground-based 
values on even the largest scales (see also Figure~\ref{fig:bigcomp}). 
Of the 24 galaxies with measured FWHM in the $1''$ aperture, the FWHM of 
only six (NGC~1358, NGC~3031, NGC~3982, NGC~4051, NGC~4151, and 
NGC~7674) are within 10\% of the ground-based value (see 
Table~\ref{tbl:width}), while 12 are discrepant by greater than 30\% -- 
comparable to the observed scatter in the $\sigma_g - \sigma_*$ correlation. 
Surprisingly, one of these 12 measurements (NGC~2110 \sii) is actually $>30$\% 
larger in the STIS aperture than in the ground-based measurement. This broad, 
nuclear component is quite obvious in Figure~\ref{fig:spectra}. For 
NGC~3516 the difference between the STIS and NW95 measurements can be 
attributed to the slit orientation as this galaxy is known to have 
a large, extended NLR \citep{Pogge89a}, yet we detect little emission 
in our slit outside of the nuclear region. 

For the seven galaxies with \oiii\ measurements, we examined the difference 
between the FWHM in our $1''$ aperture and the ground-based value 
as a function of the fraction of the flux within the $1''$ STIS aperture. 
The STIS flux measurement for NGC~7674 includes 90\%
of the ground-based value reported by \citet{Whittle92a} and our measured 
FWHM is within 10\% of his value. This confirms that when 
most of the ground-based flux falls within our much narrower aperture we 
measure the same kinematics. 
For the other six galaxies these data miss a larger fraction of the total 
flux and the line widths are quite different in some instances. The most 
notable is the STIS measurement of MKN~348, which includes 60\% of the 
ground-based value and is 60\% narrower. 
The STIS measurements of the remaining five galaxies include only 30\% of 
the flux in the \citet{Whittle92a} aperture and the differences in width 
between the two aperture sizes range between none (NGC~4151) and 
150\% (NGC~5929).

The disagreement between widths measured in our $1''$ aperture and the 
ground-based aperture is more striking for IPV20 (see also 
Table~\ref{tbl:ipv}) than for line widths as only two galaxies (NGC~4051 and 
NGC~4151) have STIS $1''$ and NW95 measurements that agree within 10\%. 
The majority (15/24) have NW95 values $>30$\% larger than the STIS measurements 
and 2/24 (NGC~7682 and NGC~2110 \oiii) are less than $30$\% of the 
ground-based measurements. 
However, we note that most differences in IPV20 between the STIS 
$1''$ and NW95 are comparisons between STIS \sii\ and NW95 \oiii\ 
measurements and they may therefore partly reflect more pronounced 
wings in the \oiii\ line relative to \sii. For example, NGC~2110 has 
a $1''$ IPV20 = 423 \kms\ for \sii\ and IPV20 = 1105 \kms\ for \oiii. NW95 
measured 655 \kms\ for \oiii. 

As described in Section~\ref{sec:asym}, there are also substantial differences 
between the fraction of AGN with asymmetries in the low-excitation \sii\ 
line in these data and ground-based measurements. This change may be due to 
the dilution of nuclear asymmetries by larger-scale emission from host galaxy 
starlight or more symmetric emission from the NLR on larger scales. 
Host galaxy dilution may in particular explain why we detect significant 
asymmetries in such a large fraction of the \sii\ profiles compared to 
expectations from ground-based programs 
\citep{Filippenko84,deRobertis84,Greene05}. 
This result implies that the asymmetries are primarily due to nuclear 
line-emitting gas, rather than material more evenly distributed throughout 
the NLR on larger scales. 

\begin{figure*}
\figurenum{7}
\plotone{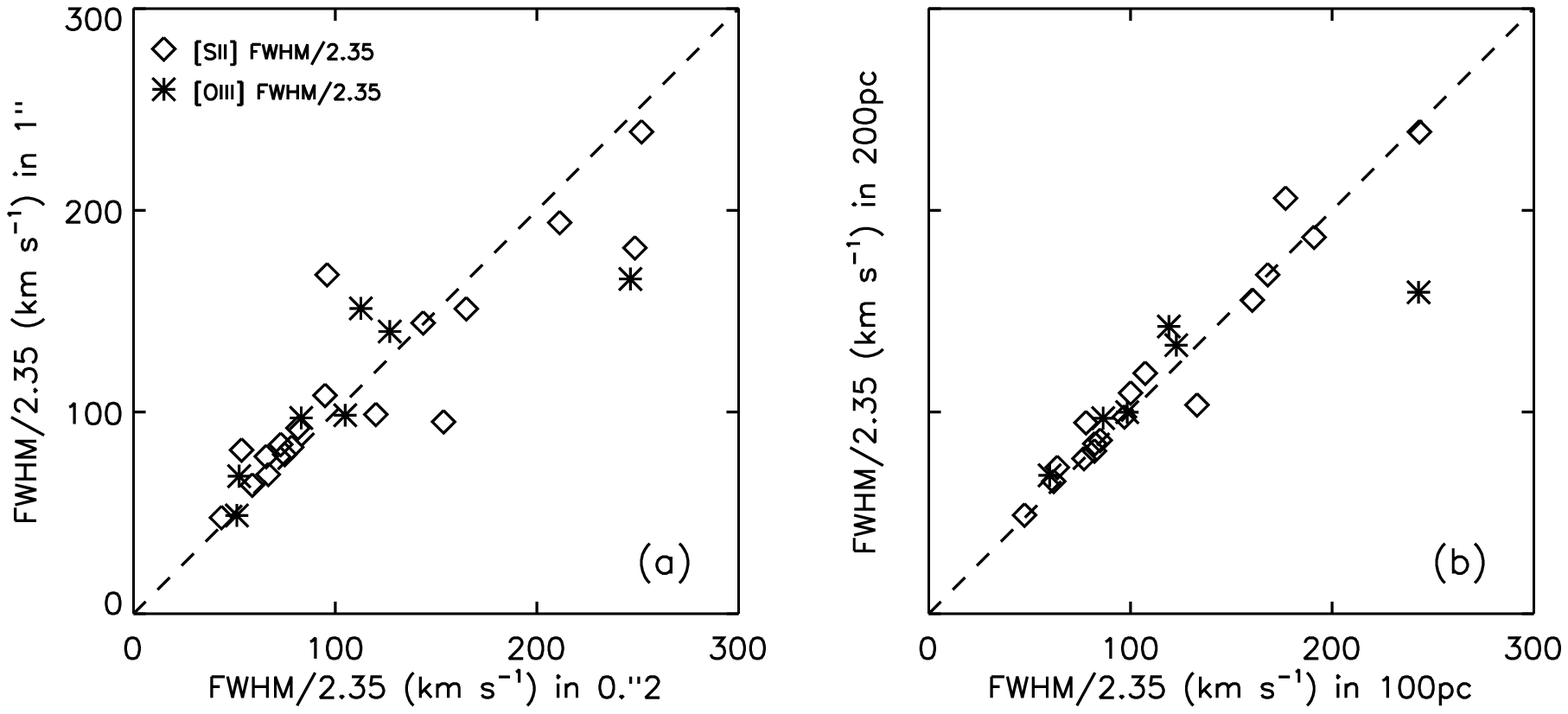} 
\caption{Comparison of our \sii\ (diamonds) and \oiii\ (stars) FWHM 
measurements in different apertures. (a) Width measured in a $1''$ aperture 
plotted against the width measured in an $0.2''$ aperture. (b) Width measured 
in a 200pc aperture plotted against the width measured in a 100pc aperture. 
The dashed line in each frame represents a 1:1 correspondence between the
quantities plotted. 
\label{fig:smallcomp} }
\end{figure*}

This comparison of width and IPV values suggest that measurements within the 
STIS aperture are generally smaller than the ground-based values. A 
quantitative comparison demonstrates that this is the case, relative to both 
ground-based NLR width measurements and $\sigma_*$ (see Table~\ref{tbl:stats}
and Figure~\ref{fig:bigcomp}). 
Comparison of the STIS FWHM measurements in $0.2''$, $1''$, 100pc, and 200pc 
apertures (divided by 2.354 to approximate a Gaussian $\sigma_g$) to $\sigma_*$ 
show that the STIS measurements systematically underestimate 
$\sigma_*$ by 10 -- 20\%. The STIS measurements similarly underestimate the 
ground-based NW95 FWHM measurements. 
Figure~\ref{fig:bigcomp} does not show any significant evidence for differences 
between \sii\ and \oiii. 
The scatter between STIS and $\sigma_*$ measurements is slightly worse 
(30 -- 40\%) than the scatter between STIS and NW95 FWHM measurements 
(20 -- 40\%). 
In contrast, Figure~\ref{fig:smallcomp} indicates that STIS measurements at 
$0.2''$ and $1''$ and 100pc and 200pc are significantly more similar in value 
and have substantially less scatter (10 -- 20\%, see also 
Table~\ref{tbl:stats}). 
The smaller measured widths in the STIS slit is likely due to the collisional 
nature of gas, which will therefore tend to reside at least partially in a 
disk. 
A narrow slit at a random orientation will then sample a smaller fraction 
of the full radial velocity field than a larger aperture slit and therefore 
measure a smaller width. 

\begin{deluxetable*}{lcccc}
\tabletypesize{\footnotesize}
\tablecaption{Statistical Results\label{tbl:stats}}
\tablenum{5}
\tablehead{
\colhead{} & 
\colhead{Aperture Size} & 
\colhead{[S~II] } & 
\colhead{[O~III] } & 
\colhead{Total} \\ 
\colhead{} & 
\colhead{} & 
\colhead{(\kms)} & 
\colhead{(\kms)} & 
\colhead{(\kms)} \\
\colhead{(1)} &
\colhead{(2)} &
\colhead{(3)} &
\colhead{(4)} &
\colhead{(5)} 
} 
\startdata
$\langle$(FWHM/2.354)/$\sigma_{*}\rangle$ & 0."2 & 0.81 $\pm$ 0.36 & 0.76 $\pm$ 0.11 & 0.80 $\pm$ 0.34 \\
 & 1" & 0.82 $\pm$ 0.31 & 0.80 $\pm$ 0.33 & 0.81 $\pm$ 0.31 \\
 & 100pc & 0.83 $\pm$ 0.31 & 0.81 $\pm$ 0.27 & 0.83 $\pm$ 0.30 \\
 & 200pc & 0.85 $\pm$ 0.30 & 0.83 $\pm$ 0.30 & 0.84 $\pm$ 0.29 \\
 &  &  &  &  \\ 
$\langle$FWHM/NW95 FWHM$\rangle$ & 0."2 & 0.84 $\pm$ 0.33 & 0.70 $\pm$ 0.34 & 0.80 $\pm$ 0.33 \\
 & 1" & 0.84 $\pm$ 0.20 & 0.70 $\pm$ 0.22 & 0.80 $\pm$ 0.21 \\
 & 100pc & 0.86 $\pm$ 0.27 & 0.72 $\pm$ 0.35 & 0.82 $\pm$ 0.29 \\
 & 200pc & 0.88 $\pm$ 0.24 & 0.70 $\pm$ 0.21 & 0.83 $\pm$ 0.24 \\
 &  &  &  &  \\ 
$\langle$FWHM/FWHM$\rangle$ & 1" / 0."2 & 1.06 $\pm$ 0.26 & 1.07 $\pm$ 0.23 & 1.06 $\pm$ 0.25 \\
 & 200pc / 100pc & 1.03 $\pm$ 0.10 & 1.04 $\pm$ 0.20 & 1.03 $\pm$ 0.13 \\
 &  &  &  &  \\
$\langle$(NW95 FWHM/2.354)/$\sigma_{*}\rangle$ &  & - - - & 1.15 $\pm$ 0.33 & - - - \\
\enddata
\tablecomments{Col. (1): Pair of values used for statistical
comparison. Col. (2): Aperture size used in our measurement of
FWHM/2.354. Col. (3): Mean and standard deviation of the sources for
which we measure the [S~II] line widths. Col. (4): Mean and standard
deviation of the sources for which we measure the [O~III] line
widths. Col. (5): Mean and standard deviation of all sources.}
\end{deluxetable*}

The comparable number of galaxies with red and blue asymmetries on larger 
scales, and more importantly that these asymmetries are in general 
substantially weaker than those in the $0.2''$ aperture, suggest that 
asymmetries measured in large (including ground-based) apertures 
originate on small scales, or less than $\sim 100$pc based on the 
spatial resolution of these observations. 

\subsection{Correlations with Global Properties} 

\begin{figure*}
\figurenum{8}
\plotone{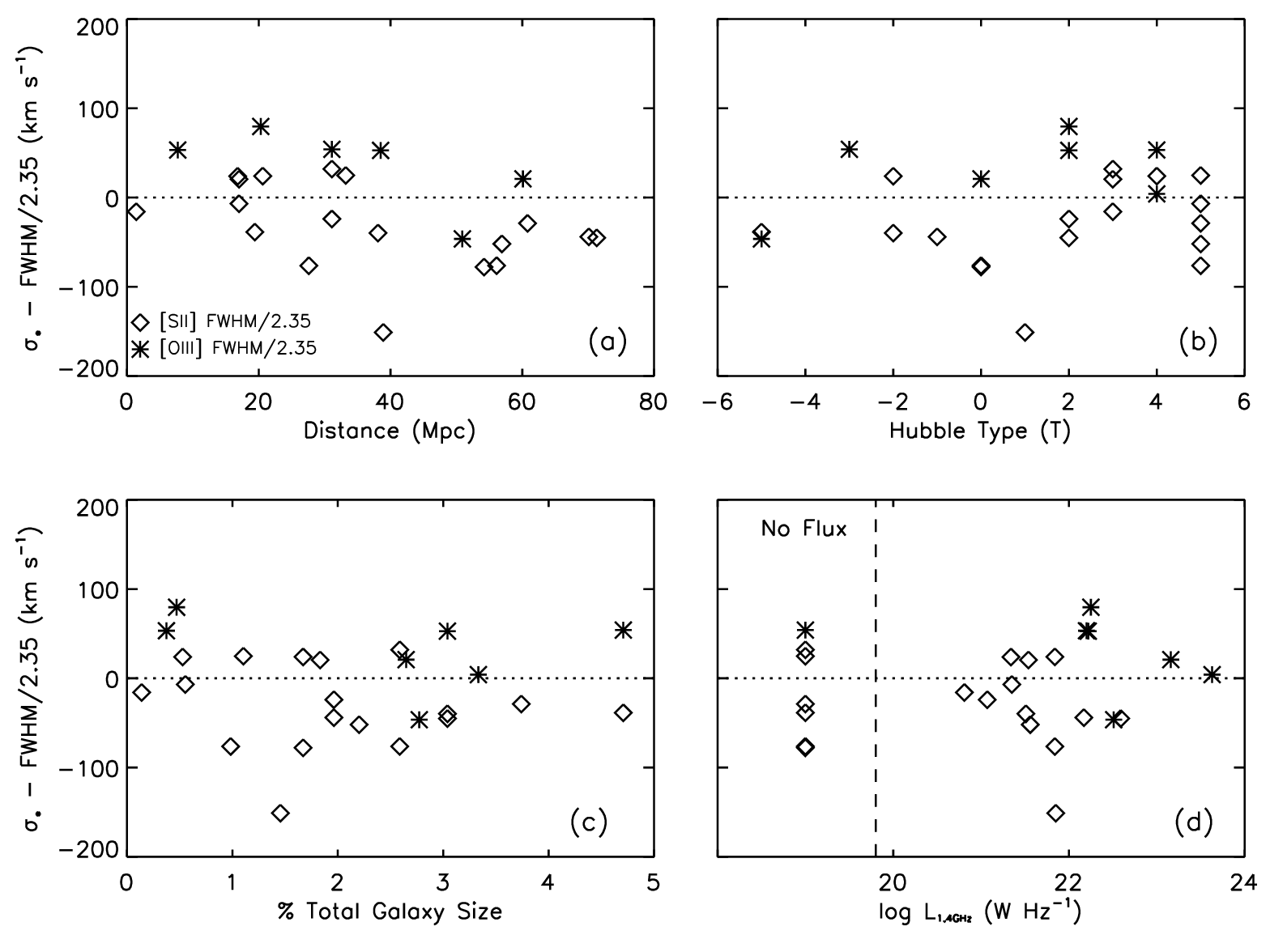} 
\caption{Comparison of residuals between gas and stellar kinematics and 
various global properties. The four panels show: (a) Distance; (b) 
Hubble T type; (c) Fraction of the galaxy size ($2''$ divided by the 
angular size provided in Table~\ref{tbl:data}); and (d) radio power. 
The points to the left of the dashed line do not have a measured radio 
flux and are only shown to indicate the ordinate values. 
\label{fig:res} }
\end{figure*}

In addition to the quality of the correlations between line widths measured 
in various apertures and $\sigma_*$, we also searched for systematic 
trends between the residuals and the properties of the galaxies. We 
specifically investigated the residuals between the measurements in the 
$1''$ aperture and $\sigma_*$ as a function of distance, Hubble type, 
fraction of the galaxy in a $1''$ aperture, and radio power. 
Significant residuals as a function of galaxy distance, Hubble type, and 
fraction of the angular size of the galaxy subtended by the spectroscopic slit 
would indicate very useful information about the origin of the scatter in the 
$\sigma_g - \sigma_*$ relation and these quantities could potentially be 
used to empirically determine corrections to reduce scatter, as well as 
provide insight into its origin. For example, the NLR measured exactly within 
the host bulge's effective radius might prove to be the best tracer of 
$\sigma_*$. Some evidence for such a relation might be revealed in 
the residuals plotted against these three parameters, however as shown in 
Figure~\ref{fig:res} there is no evidence that the residuals are correlated 
with any of these parameters. 

\citet{Whittle92c} found evidence that Seyfert galaxies with linear 
radio morphology and high luminosity tend to have broader lines. 
We do not see significant correlation between residuals in 
$\sigma_g - \sigma_*$ in our data, although only three galaxies in our 
sample are above his radio luminosity threshold of 
$L_{1415} \geq 10^{22.5}$ W Hz$^{-1}$. 
Other parameters mentioned in the Introduction that correlate with 
systematically broader emission-line widths include $L/L_{Edd}$ 
\citep{Greene05} and disturbed morphologies \citep{Whittle92b}, however this 
sample does not contain a sufficient number of objects with either high 
accretion-rates or significantly-disturbed morphologies to investigate 
these quantities. \citet{Boroson05} also recently noted that strongly 
blueshifted \oiii\ lines (relative to the host galaxy) were systematically 
broader than similar objects that were not blueshifted. These objects 
tended to be those with large $L/L_{Edd}$. None of the seven galaxies 
in our \oiii\ sample have substantial blueshifts. 

\section{Summary}  \label{sec:sum}

We have conducted a detailed analysis of spatially-resolved \oiii\ and \sii\ 
NLR emission from 24 well-studied, nearby AGN. 
These observations have detected considerable emission outside of the 
unresolved nucleus ($0.2''$ or 10 -- 100pc) and this emission often 
contributes significantly to the measured line profiles at larger scales. 
We have characterized the spatial dependence of this emission with a range of 
width, area, and asymmetry measurements and shown there are not only 
substantial changes from $0.2''$ to $1''$ (typical of ground-based 
observations), but also from a $1''$ long STIS aperture (with either a 
$0.1''$ or $0.2''$ width) and the approximately $1'' \times 1''$ or 
larger aperture used for ground-based measurements of these galaxies. 
In particular, the large variations in line profiles as a function of 
aperture size demonstrates that the profile of the NLR lines are set by 
the kinematics of the gas in the NLR itself, and not just radiatively-driven 
winds from the nucleus. 

The spatial scale(s) responsible for the NLR emission lines has received 
substantial recent attention because the widths of the NLR lines may 
be used to estimate the stellar velocity dispersion $\sigma_*$, which 
in turn can be used as a proxy for the black hole mass $\mbh$. Our analysis 
shows that even at the largest spatial scales observed with STIS, 
the line widths are systematically smaller by 10 -- 20\% than 
ground-based width measurements (as well as $\sigma_*$). As we estimate that 
the STIS slit may include as little as 25\% of the NLR flux, emission on larger 
scales or outside of the narrow STIS slit width must broaden the line 
profiles. 

While the line profile measurements from the STIS data are systematically 
less than the ground-based NLR and $\sigma_*$ values, the scatter is 
comparable. The substantial scatter in the $\sigma_g - \sigma_*$ relation 
therefore appears to be due to a sufficiently complex set of parameters that 
it is effectively stochastic, at least to the extent that the scatter can not 
be reduced to the magnitude of the scatter in the $\mbh - \sigma_*$ 
relation. In addition to rotation, Eddington ratio, and compact radio jets, 
the list of parameters that increase the scatter should include the clumpiness 
of the NLR, the orientation of the NLR with respect to the host galaxy 
semimajor axis, and the amount and distribution of dust in the NLR. 
If the NLR were as isotropic as the stellar distribution, the scatter 
would be comparable to the $\mbh - \sigma_*$ relation, yet because the gas 
is collisional, and also clumpy, it can not be as good a tracer 
of the bulge potential. 

This interpretation is supported by the substantial and diverse types of width 
variations observed in these spatially-resolved measurements. 
Specifically, the most common types of line width variation are $<30$\% 
changes between the $0.2''$, $1''$, and NW95 apertures (nine galaxies) and 
relatively little width increase in the STIS aperture, but a substantial 
increase ($>30$\%) between the $1''$ STIS and NW95 measurements (also nine 
galaxies). 
The remaining types of variations observed are greater than 30\% width 
changes on the STIS scales only (four) and on all scales (two). 
Two of the four galaxies with only substantial profiles changes on the STIS 
scale have substantially broader emission in the $0.2''$ than in the $1''$ 
aperture. These large nuclear velocities may be due to radiatively-driven 
winds that decelerate or simply do not extend to larger scales. However, 
neither of these nuclear line profiles exhibit substantial asymmetries, 
which suggests that if this emission is due to a uniform outflow, there is not 
substantial dust in the nuclear region. 

Both red and blue asymmetries are observed in the \sii\ and \oiii\ line 
profiles. These observations are unusual in two respects. First, asymmetries 
in \sii\ are reported far less frequently than in \oiii, yet we observe 
\sii\ asymmetries with nearly comparable frequency. The low frequency of 
observed \sii\ asymmetries in ground-based observations have been ascribed 
to the fact that \sii\ is a lower-ionization line. Our observations suggest 
that asymmetries are commonly present in \sii, but the weaker \sii\ emission
is more easily diluted by host galaxy starlight and not as readily 
observed in the larger ground-based apertures. Secondly, we observe several 
examples of galaxies with red asymmetries in either \sii\ or \oiii. For 
the cases in which the red asymmetries are largely confined to the unresolved 
nuclear spectrum, these observations do not agree well with outflow models 
that produce blue asymmetries through invocation of obscuration of the 
far side of the galaxy. These red asymmetries may be due to the patchy nature 
of the ISM on very small scales, which may produce uneven illumination of 
NLR clouds and a patchy line-of-sight velocity distribution for the NLR 
clouds. Such variations within the NLR may be responsible for the bulk 
of the scatter in the $\sigma_g - \sigma_*$ correlation.

\acknowledgements

We thank the referee for helpful comments on the manuscript. 
Support for this work was provided by NASA through grant numbers
AR-9547 and GO-9143 from the Space Telescope Science Institute,
which is operated by the Association of Universities for Research in 
Astronomy, Inc., under NASA contract NAS5-26555.   
PM was supported in part by a Clay Fellowship from the Harvard-Smithsonian 
Center for Astrophysics.

%%% FIGURES 

\end{document}